\documentstyle[12pt]{article}
\input mssymb

\newcommand{\uu}[1]{\verb!#1!\endgroup}     

\newcommand{\mb}[1]{\ifmmode#1\else{\ifinner#1\else\mbox{$#1$ }\fi}\fi}

\newcommand\al{\mb{\alpha}}

\newcommand\be{\mb{\beta}}
\newcommand\ga{\mb{\gamma}}
\newcommand\de{\mb{\delta}}
\newcommand\ep{\mb{\epsilon}}
\newcommand\varep{\mb{\varepsilon}}

\newcommand\et{\mb{\eta}}
\newcommand\th{\mb{\theta}}

\newcommand\ka{\mb{\kappa}}
\newcommand\la{\mb{\lambda}}
\newcommand\muu{\mb{\mu}} 
\newcommand\nuu{\mb{\nu}} 
\newcommand\xii{\mb{\xi}}
\newcommand\pii{\mb{\pi}}
\newcommand\rh{\mb{\rho}}
\newcommand\si{\mb{\sigma}}
\newcommand\ta{\mb{\tau}}

\newcommand\ph{\mb{\phi}}

\newcommand\ch{\mb{\chi}}
\newcommand\ps{\mb{\psi}}
\newcommand\om{\mb{\omega}}
\newcommand\Ga{\mb{\Gamma}}
\newcommand\De{\mb{\Delta}}
\newcommand\Th{\mb{\Theta}}
\newcommand\La{\mb{\Lambda}}

\newcommand\Ph{\mb{\Phi}}
\newcommand\Ps{\mb{\Psi}}
\newcommand\Om{\mb{\Omega}}

\newcommand\calA{\mb{{\cal A}}}
\newcommand\calB{\mb{{\cal B}}}

\newcommand\calF{\mb{{\cal F}}}
\newcommand\calG{\mb{{\cal G}}}
\newcommand\calH{\mb{{\cal H}}}

\newcommand\calL{\mb{{\cal L}}}

\newcommand\calO{\mb{{\cal O}}}

\newcommand{\eq}[1]{\begin{equation}#1\end{equation}}
\newcommand{\eqn}[1]{\begin{eqnarray}#1\end{eqnarray}}

\font\bigm=cmex10 scaled\magstep1
\def\bigint_#1^#2{\hbox{\raise16.8pt\hbox{\bigm\char'132}}_{#1}^{\;\;#2}}

\font\Bigm=cmex10 scaled\magstep2
\def\Bigint_#1^#2{\hbox{\raise19.4pt\hbox{\Bigm\char'132}}_{\!#1}^{\:\;\;#2}}

\font\biggm=cmex10 scaled\magstep3
\def\biggint_#1^#2{\hbox{\raise23.2pt\hbox{\biggm\char'132}}_{\!#1}^{\;\;\;#2}}

\font\Biggm=cmex10 scaled\magstep4
\def\Biggint_#1^#2
{\hbox{\raise28.4pt\hbox{\Biggm\char'132}}_{\!\!#1}^{\:\;\;\;#2}}

\font\BIGm=cmex10 scaled\magstep5
\def\BIGint_#1^#2{\hbox{
\raise33.5pt\hbox{\BIGm\char'132}}_{\!\!\!#1}^{\:\;\;\;\;#2}}

\font\normalm=cmex10
\def\normalint_#1^#2{\hbox{\raise15pt\hbox{\normalm\char'132}}_{#1}^{\;\;#2}}

\def\smallint_#1^#2{\hbox{\raise10pt\hbox{\bigm\char'122}}_{#1}^{\;#2}}

\newdimen\mycirclesize
\newcount\mycirclesiz

\def\mycircl#1{%
\mycirclesiz\mycirclesize%
\mycirc{\the\mycirclesiz}%
}
\def\mycirc#1{%
\special{" newpath 0 0 #1 65536 div 0 360 arc stroke }%
}
\newcommand{\cinf}[1]{\mb{C^{\infty}(#1)}}
\newcommand{\x}{\mb{\times}}

\newcommand{\R}{\mb{\Bbb R}}
\newcommand{\hst}{\hspace{10mm}}
\newcommand{\fa}{\forall}
\newcommand{\TS}[1]{\mb{T^{*}#1}}

\newcommand{\w}{\mb{\wedge}}
\newcommand{\cd}{\mb{\partial}}

\newcommand{\g}{\mb{{\bf g}}}
\newcommand{\gs}{\mb{{\bf g}^*}}
\newcommand{\gsp}{\mb{{\bf g}^*_+}}
\newcommand{\gsm}{\mb{{\bf g}^*_-}}

\newcommand{\h}{\mb{{\bf h}}}
\newcommand{\hs}{\mb{{\bf h}^*}}
\newcommand{\m}{\mb{{\bf m}}}
\newcommand{\ms}{\mb{{\bf m}^*}}

\newcommand{\iso}{\simeq}
\newcommand{\cdd}[2]{{\mb{{\cd #1}\over{\cd #2}}}}
\newcommand{\co}{\mb{\pi_{co}}}

\newcommand{\eo}{\setcounter{equation}{0}}
\newcommand{\beq}{\begin{equation}}
\newcommand{\eeq}{\end{equation}}
\newcommand{\nn}{\nonumber}
\newcommand{\crc}{\mb{{\scriptstyle\circ}}}
\newcommand{\ssect}{\eo\subsection} 
\newcommand{\sssect}{\subsubsection} 

\newcommand{\bea}{\begin{eqnarray}}
\newcommand{\eea}{\end{eqnarray}}
\newcommand{\emfl}[1]{\vspace{\baselineskip}\noindent{\em#1}\par}
\newcommand{\JH}{\mb{\hat J}}
\newcommand{\cod}[1]{\mb{\co(#1)\cdot}}
\newcommand{\rst}{\mb{\restriction}}
\newcommand{\lx}{\mb{\ltimes}}

\newcommand{\su}{\mb{{\rm\bf su(2)}}}
\newcommand{\so}{\mb{{\rm\bf so(3)}}}

\newcommand{\del}{\mb{\nabla}}

\newcommand{\hbm}{\mb{\hbar^{-1}}}
\newcommand{\bs}{\backslash}

\newcommand{\ns}{\mb{{\bf n}^*}}
\def\one{{1\kern -.60em 1}}
\newcommand{\bt}{\mb{{\bf t}}}
\newcommand{\bb}{\mb{{\bf b}}}
\renewcommand{\i}{\mb{\imath}}
\newcommand{\ci}{\cite}
\newcommand{\sect}{\section}
\renewcommand{\ll}{\label}

\topmargin = - 0.5 cm
\newtheorem{theorem}{Theorem}
\newtheorem{lemma}{Lemma}

\jot
\begin{document}
\bibliographystyle{plain}

\begin{flushright}
DAMTP 94-36
\end{flushright}
\begin{center}
\LARGE{Geometric Quantization of the Phase Space of a Particle in a
Yang-Mills Field\\}
\vspace*{0.5cm}
\large{M.\ A.\ Robson\footnote{E-mail:
M.A.Robson\symbol{64}amtp.cam.ac.uk.}\footnote{Supported by an
E.P.S.R.C.\ studentship.}\\ }
\vspace*{0.2cm}
{\small\em Department of Applied Mathematics and Theoretical
Physics,\\
University of Cambridge, Silver Street,\\
Cambridge, CB3 9EW, U.\ K.\\ }
\vspace*{0.2cm}

{May 1994}
\end{center}

\begin{abstract}
The method of geometric quantization is applied to a particle moving
on an arbitrary Riemannian manifold $Q$ in an external gauge field,
that is a connection on a principal $H$-bundle $N$ over $Q$. The phase
space of the particle is a Marsden-Weinstein reduction of $T^*N$, hence
this space can also be considered to be the reduced phase space of a
particular type of constrained mechanical system. An explicit map is
found from a subalgebra of the classical observables to the
corresponding quantum operators. These operators are found to be the
generators of a representation of the semi-direct product group,
Aut~$N\lx C^\infty_c(Q)$. A generalised Aharanov-Bohm effect is shown
to be a natural consequence of the quantization procedure. In
particular the r\^ole of the connection in the quantum mechanical
system is made clear. The quantization of the Hamiltonian is also
considered.

Additionally, our approach allows the related quantization procedures
proposed by Mackey and by Isham to be fully understood.\\
\\
\parbox[t]{2cm}{\em Keywords:}
\parbox[t]{10.77cm}{\em geometric quantization, Marsden-Weinstein
reduction, constrained systems, external gauge fields}\\
\parbox[t]{2cm}{\em MSC:}
\parbox[t]{10.5cm}{\em 81 S 10, 58 F 06}\\
\parbox[t]{2cm}{\em PACS:}
\parbox[t]{10.5cm}{\em 03.65Bz}
\end{abstract}
\thispagestyle{empty}
\newpage
\setcounter{page}{1}
\sect{Introduction}
\ll{intro}
Constrained mechanical systems make up an important category of
classical dynamical systems. We consider the case where the
constrained system is described by a symplectic manifold $S$ (the
``unconstrained'' phase space of the system) together with a Lie group
$H$, which has a Hamiltonian action on $S$, and a corresponding
equivariant momentum map $J_H:S\to\hs$ where \h is the Lie algebra of
$H$ and \hs is its dual. The constraints are given by $J_H=\mu$ for
some fixed $\mu\in\hs$. Subject to certain technical assumptions, the
reduced phase space of the system (i.e., the true phase space of the
system in which the constraints are automatically satisfied) is then a
quotient manifold of $J^{-1}(\mu)$ and inherits a symplectic structure
from $S$ \ci{MW}; this particular method of identifying the reduced
phase space is called Marsden-Weinstein reduction. A key point is that
this quotient manifold is symplectomorphic to a symplectic leaf in the
Poisson manifold $S/H$ \ci{Marle,KKS}. The Poisson bracket on $S$
drops to one on $S/H$ (since $H$ acts symplectically on $S$) and this
defines the Poisson structure on $S/H$.

We concentrate on the case where $S$ is a cotangent bundle $T^*N$. We
assume that $H$ acts freely on $N$ so that we may consider
$(N,Q,H,\pi_{N\to Q})$ to be a principal fibre bundle with total space
$N$, base space $Q=N/H$, projection $\pi_{N\to Q}$, and where the Lie
group $H$ acts on the right of $N$. Thus $H$ acts on $S=T^*N$ by
cotangent lift and there is always an equivariant momentum map for
this action
\ci{AM}. The reduced phase space is then a symplectic leaf in
$(T^*N)/H$.

One important physical interpretation, originally due to Sternberg, of
this type of constrained system, is well known in the context of a
charged particle moving on $Q$ in the presence of an external
Yang-Mills field with gauge group $H$
\ci{Stern,Wein78,Mont,GS}. Specifically $S/H=(T^*N)/H$ is the
``universal phase space'' of the particle. There is a one-to-one
correspondence between the symplectic leaves of $S/H$ and the
coadjoint orbits in $\hs$. Each of the latter represents a different
possible charge of the particle so that $S/H$, which is foliated by
its symplectic leaves, is composed of the phase spaces corresponding
to every possible charge. In respect to this example we often refer to
$Q$ as the configuration space and $H$ as the gauge group.

Naturally the construction of the quantum mechanical system
corresponding to a constrained mechanical system has aroused much interest.
Recall that quantization tries to associate to each classical system
(described by a symplectic manifold $S$) a Hilbert space, $\calH$, of
quantum states and a map from the space of classical observables
(smooth functions on $S$) to the space of symmetric operators,
$\calO$, on $\calH$. Each classical observable $f\in\cinf S$ should
correspond to an operator $\hat f\in\calO$ such that \begin{description}
\item[(Qi)] the map $f\to\hat f$ is linear (over $\Bbb R$);
\item[(Qii)] if $f=\one_S$, then $\hat f=\one_\calH$, where \one\ denotes
the identity operator;
\item[(Qiii)] if $\{f_1,f_2\}=f_3$ then $[\hat f_1,\hat f_2]=i\hbar
\hat f_3$.
\end{description}
Additionally, some sort of irreducibility condition is also imposed.
When $S=T^*Q$ is a cotangent bundle the operators $\hat q$ and $\hat
p$ corresponding to $(q,p)\in T^*Q$ are required to act irreducibly
whilst when $S$ is a coadjoint orbit the map $f\to \hat f$ must give an
irreducible representation of the generators of the symmetry group. In
order to meet this requirement of irreducibility, restrictions are
imposed, in all quantization schemes, on the class of observables that
can be quantized. However, for a {\em general} symplectic manifold
there appears to be, in the literature, no definite statement of the
irreducibility requirements. We shall find though, that our method
associates with quantization a representation of a Lie group which is,
in general, irreducible, thus meeting any reasonable irreducibility
requirement.

Much work \ci{Gotay,Puta,BRST,nonuni,Blau} has been done on the
quantization of the reduced phase spaces of constrained systems and of
a particle in a gauge field, most notably by Landsman for the case of
homogeneous configuration spaces
\ci{Landhom} using induced representations and for the general case
\ci{LandSD,LandR} using Rieffel's notion of ``strict deformation
quantization'' and Rieffel induction respectively. However, these
approaches suffer from the inability, in general, to quantize
classical observables which are unbounded. The method of geometric
quantization has so far been restricted to comparing the quantization
obtained by first solving the constraints (i.e., reduction) and then
quantizing the reduced space or quantizing the extended phase space
and then imposing the constraints at a quantum level. However, in the
setting that we work in, it was found
\ci{Gotay} that the two approaches were equivalent only if the reduced
phase space was symplectomorphic to a cotangent bundle and thus the
 geometric quantization of the reduced phase space was restricted
to this rather special case.

Another notable contribution is in the area of homogeneous
configuration spaces $Q$. This study was initiated by Mackey \ci{Mac}
and was extended by Isham \ci{Isham}, who used a group-theoretic
approach to identify a particular semi-direct product group, $\calG$,
which acted on the phase space $T^*Q$ of the system.  Quantization
then corresponded to assigning quantum operators to be generators of
an irreducible unitary representation of the group $\calG$.  However,
as in general, there is more than one such representation of this
group, many different inequivalent quantum systems arise from the
study of the same configuration spaces. We will see that these
correspond to the geometric quantization of different symplectic
leaves of $(T^*G)/H$ where $G$ is a Lie group ($H\subset G$) so that
$G/H$ is the homogeneous space $Q$. (Note that $T^*Q\subset
(T^*G)/H$.) Indeed, the underlying motivation for this paper was the
anticipation of this result, which was based upon two previously known
results. Firstly, it has been shown \ci{MRW} that the symplectic
leaves of $(T^*G)/H$ are symplectomorphic to certain coadjoint orbits
in the dual of the Lie algebra of $\calG$. Whilst secondly, Rawnsley
\ci{Rawn} has shown that the geometric quantization of these orbits
leads to the induced representations upon which Mackey theory is
based.

One important feature of Isham's approach
was the use of a momentum map to relate the classical observables with
their quantum operator counterparts. Specifically if
\calG is a Lie group, with Lie algebra $\calL(\calG)$, which acts on
the symplectic manifold $M$ and $J:M\to\calL(\calG)^*$ is a
corresponding equivariant momentum map then, given a representation $\pi$ of
$\calG$, a ``quantizing map'', $Q_\hbar$, can be given which relates
classical observables to quantum operators. Explicitly,
\eq{Q_\hbar(\hat J(A))=\hbar d\pi(A),\ll{qzmap}}
where $\hat J:\calL(\calG)\to\cinf M$ is defined by $\langle
J(m),A\rangle=(\hat J(A))(m)$ and $d\pi$ is the derived Lie algebra
representation, where we are following the convention that
\eq{d\pi(A)=i\left.\frac{d}{dt}\pi(e^{tA})\right|_{t=0}.\ll{derivdef}}
The general properties of a momentum map then ensure that condition
(Qiii) above is automatically satisfied. The subclass of observables
that can be quantized by $Q_\hbar$ is clearly $\{\hat J(A):A\in
\calL(\calG)\}$. Clearly, this approach hinges on the ability to choose
\calG and \pii correctly.

This paper uses the geometric quantization framework of Kostant and
Souriau to give a complete quantization of the constrained mechanical
system whose reduced phase space is [symplectomorphic to] a symplectic
leaf in $(T^*N)/H$. In particular, our only assumptions are that the
space $N$ has a Riemannian structure with an $H$-invariant metric (so
that $Q$ inherits a metric from $N$) and that the gauge group $H$ is a
connected and compact Lie group. We are able to combine naturally the
group-theoretic and geometric quantization approaches, finding on the
way how each sheds light on the other. In particular, we are able to
present our results in the language of representations so that the
quantum operators are given as generators of a representation of a Lie
group, together with a corresponding momentum map which explicitly
links the quantum operators with their classical observable
counterparts in the manner described above. Thus, no knowledge of
geometric quantization is required in order to appreciate the results
found.

Our presentation relies very heavily on the combination of the
symplectic formulation of constrained mechanical systems with the
method of geometric quantization. Since no one source adequately
presents both theories in the detail and manner needed, a
short review of both is given in sections \ref{generalsym} and
\ref{gq} respectively. In particular, the final subsection of section
\ref{generalsym} gives a new result regarding the action
of the semi-direct product group, $\calG={\rm Aut\ }N\lx\cinf Q$, of
fibre preserving diffeomorphisms of $N$ and smooth functions on $Q$ on
the phase space of the reduced system. This group action has a
corresponding momentum map and the idea is to quantize in the style of
(\ref{qzmap}). Indeed section \ref{gqgen} can be regarded essentially
as justifying this choice of \calG and showing which representation
\pii of \calG is to be chosen in the right hand side of
(\ref{qzmap}). Subsection
\ref{homspaces} explicitly compares our approach with that of Isham's
\ci{Isham} for homogeneous spaces.

Section \ref{gqreview} reviews very briefly the method of geometric
quantization and ends with subsection \ref{Hqz} where a known result,
concerning the induced representation found from [geometrically]
quantizing a coadjoint orbit, is restated with slightly weaker
conditions. Finally, section \ref{gqgen} forms the heart of the
paper. Using geometric quantization the reduced phase space is
quantized. We find that, using a particular polarization, the subclass
of observables that can be quantized is the same as that predicted by
use of the group $\calG={\rm Aut\ }N\lx\cinf Q$. We then show that the
corresponding quantum operators are generators of a representation
\pii of $\calG$, the choice of \pii depending on which symplectic leaf
of $(T^*N)/H$ we are quantizing on. Along the way we find that the
Aharanov-Bohm effect is a natural consequence of our quantization and
also that the nonintegrable phase factor of Wu and Yang \ci{WY}
appears in the analogous result for the case when the gauge
group is non-Abelian. Using the results of
\ci{LandSD} we are able to give a Hamiltonian for the quantum system
which then completes the quantization of the constrained system.

\ssect*{Acknowledgements}
The author would like to greatly thank N.\ P.\ Landsman for not only
suggesting this line of research but also for his many helpful
comments and suggestions. Additionally, N.\ Linden is thanked for
several useful discussions.
\sect{Constrained Mechanical Systems}
\ll{generalsym}
\ssect{Dual pairs and momentum maps}
\ll{symleaves}
\ll{dualpairs}
We start by reviewing the basic ideas of dual pairs and
momentum maps which provide great insight into the theory of
Marsden-Weinstein reduction. The main references for this subsection
are Weinstein \cite{Wein}, Choquet-Bruhat {\em et al} \cite[chapter
12]{AMPII} and Abraham and Marsden \ci{AM}.

A useful idea is the notion of a {\em realisation} of a Poisson
manifold $M$. This is a symplectic
manifold $S$ together with a {\em Poisson map} $J:S\to M$. A Poisson map is
one which preserves the Poisson bracket, i.e.,
\[\{J^*F,J^*G\}_{_S}=J^*\{F,G\}_{_M}\hst\fa F,G\in\cinf{M}.\]
We are interested in the case when the fibres $J^{-1}(m),\,m\in M$, define a
foliation \Ph of $S$ in such a way that $S/\Ph$ is a manifold and so
if $\pi:S\to S/\Ph$ is the canonical projection, the space
$\pi^*\cinf{S/\Ph}$ is a Lie sub-algebra of $\cinf{S}$ and coincides
with $J^*\cinf{M}$. We
denote the functions constant on the leaves of \Ph by $\calF_\Ph$ and
the functions
which Poisson commute with all elements of $\calF_\Ph$ by $\calF_{\Ph^\perp}$,
i.e., symbolically $\{\calF_\Ph,\calF_{\Phi^\perp}\}=0$. It can be shown
that this defines
a foliation $\Ph^\perp$ of $S$ such that $S/\Ph^{\perp}$ is a manifold and
the $\calF_{\Ph^\perp}$ are functions constant on the leaves of $\Ph^{\perp}$.
We call $\calF_\Ph$ and $\calF_{\Ph^\perp}$ {\em polar} to each
other.

A {\em dual pair} is where we have two Poisson manifolds $M_1$ and $M_2$ and a
symplectic manifold $S$ with Poisson maps $J_1,J_2$ between $S$ and each $M_i$
\[M_2\stackrel{J_2}\gets S\stackrel{J_1}\to M_1\]
and $\calF_{\Phi_1}$ and $\calF_{\Phi_2}$ are polar to each other.
The dual pair is called {\em full} if $J_1$ and $J_2$ are both submersions.
However, if $J_1$ and $J_2$ have constant rank then $J_1(S)$ and $J_2(S)$
are Poisson submanifolds and
$J_2(S)\stackrel{J_2}\gets S\stackrel{J_1}\to J_1(S)$
is a full dual pair. Assuming this to be the case, then
the key result is that we can define a bijection between the symplectic
leaves of $M_1$ and $M_2$ (assuming that $J_1$ and $J_2$ have
connected fibres). Specifically, if
$J_1^{-1}(m)$ is connected, then $J_2(J_1^{-1}(m))$
is a symplectic
leaf of $M_2$. In general, $J_2(J_1^{-1}(m))$ will be a union of
[connected] symplectic leaves of $M_2$.

\ll{momentummaps}
For the case where a Lie group $G$ acts symplectically on [the left
of] a symplectic manifold $S$ (i.e., the Poisson bracket is invariant
under the action of $G$) such that $S/G$ is a manifold we can often
find a {\em momentum map} $J$ such that we have the dual pair
\eq{\gsp\stackrel{J}\gets S\stackrel{\pi^G}\to S/G,}
where $\pii^G$ is the projection map $S\to S/G$ and \gsp is the dual of
the Lie algebra of $G$ with the ``+'' Lie-Poisson structure. This
means we can find the symplectic leaves of $S/G$ using the above result.

Recall that the function groups corresponding to $J$ and $\pi^G$ must be
polar to each other. This gives us our first condition on $J$.
Denote the infinitesimal generator of the left action of $G$
on $S$ by $\xii$, i.e.,
\eq{(\xi(X)f)(m)={d\over
dt}f(e^{tX}\cdot m)\Bigl\arrowvert_{t=0},\hst X\in\g,\,m\in
S,\,f\in\cinf{S}\ll{xidef},}
where we are denoting the action of $x\in G$ on $m\in S$ by $x\cdot
m$. (Note that, if $G$ acts on the right on $S$, then $e^{tX}\cdot m$
in (\ref{xidef}) is replaced by $m\cdot e^{tX}$ and \gs now has the
``$-$'' Lie-Poisson structure.) Now
define $\JH:\g\to\cinf{S}$ to be the restriction of $J^*$ from
$\cinf{\gs}$ to $\g$ (regarding $\g\subset\cinf{\gs}$, i.e.,
$X(\th)\equiv \langle \th,X\rangle $ for $X\in\g$ and $\th\in\gs$). So
explicitly
\eq{\JH(X)(m)=\langle J(m),X\rangle \ll{Jhatdef}.}
Then the first condition is that \JH must satisfy
\eq{\{f,\JH(X)\}=\xi(X)f\hst\fa f\in\cinf{S},\,\fa X\in\g\ll{Jcond}.}
Note that we define the Hamiltonian vector field, $\xi_f$, of
$f\in\cinf S$ by $\xi_fg=\{g,f\}$ for $g\in\cinf S$, so that
(\ref{Jcond}) can be written $\xi_{\hat J(X)}=\xi(X)$.
Secondly $J$ must also be a Poisson map; this
is achieved by the condition that $J$ must be equivariant, i.e.,
\eq{J(x\cdot m)=\cod{x}J(m).\ll{equivariant}}
Here $\pi_{co}(x)\equiv Ad^*_{x^{-1}}$ denotes the coadjoint action.
This last condition implies that
\eq{\{\JH(X),\JH(Y)\}=\JH([X,Y])\ll{Jcond2},}
 i.e., \JH preserves the Lie algebra structure. With these two
conditions it can be shown that we have a dual pair as described.

Now, assuming that $G$ is connected, the symplectic leaves of \gs are
coadjoint orbits. Also note that (\ref{equivariant}) implies that
$\pi^G( J^{-1}(\cod{g}\mu))=\pi^G (J^{-1}(\mu))$. Thus, assuming
the fibres $J^{-1}(\mu)$ are connected, the symplectic leaves of $S/G$
can be written $P_\calO=(J^{-1}(\calO_\mu))/G$ where $\calO_\mu$ is a
coadjoint orbit in $\gs$. The symplectic form $\Om_\calO$ on $P_\calO$
is given, e.g., \cite{Mars}, via
\eq{j^*_\calO\Om=pr^*\Om_\calO+J^*_\calO\om_\calO\ll{symformdiff},}
where $j_\calO:J^{-1}(\calO)\to S$ is the inclusion, \Om is the
symplectic form on $S$,
$pr:J^{-1}(\calO)\to P_\calO$ is the projection $\pi^G$ acting on
$J^{-1}(\calO)$; and where $J_\calO=J\rst J^{-1}(\calO):J^{-1}(\calO)\to
\calO$ and  $\om_\calO$ is the symplectic form on the coadjoint orbit.

There is an alternative expression for the symplectic leaves of $S/G$.
If $G_\mu$ denotes the {\em isotropy group} of $\mu$, i.e.,
\eq{G_\mu=\{g\in G:\cod{g}\mu=\mu\}\ll{isotgp},}
then the symplectic leaf $P_\calO=\pi^G(J^{-1}(\mu))\iso
J^{-1}(\mu)/G_\mu=P_\mu$. This process of identifying $P_\mu$ or
$P_{\calO_\mu}$ is called Marsden-Weinstein reduction.
The symplectic structure, $\Om_\mu$, on
$P_\mu$ is given by $i^*_\mu\Om=\pi^*_\mu\Om_\mu$, where
$i_\mu:J^{-1}(\mu)\to P_\mu$ is the inclusion map and
$\pi_\mu:J^{-1}(\mu)\to P_\mu$ is the projection map.

There is an important general result regarding the actions of a group
$G$ on a manifold $Q$. Specifically the induced action of $G$ on the
cotangent bundle $T^*Q$ (by cotangent lift) is symplectic with respect
to the canonical symplectic form $\si_0=-d\th_0$ on $T^*Q$. Here
$\th_0$ is the canonical one-form defined by
\eq{\langle\th_0,v\rangle_{\be_q}=\langle\be_q,\pi_*v\rangle_q\hst\fa
v\in T_{\be_q}(T^*Q),\ll{canoneform}}
where $\be_q\in T^*_qQ$ and $\pi:T^*Q\to Q$ is the canonical projection.
An equivariant momentum map for the action of $G$ on $T^*Q$ is
given by $J:T^*Q\to\gs$ with
\eq{\langle J(p_q),X\rangle =\langle p_q,\xi(X)\rangle ,\hst
X\in\g.\ll{genmommap}}

There is a slightly different approach, at least in the language used,
to finding the symplectic leaves of $S/G$. In this terminology, e.g.,
\cite{Wood}, the submanifold $J^{-1}(\mu)\subset S$ is called a {\em
presymplectic manifold}. It has a two-form, $\si'$, given by just
restricting the symplectic form on $S$ to $J^{-1}(\mu)$. The {\em
characteristic distribution} of $\si'$ is
\eq{K_m=\{X:i_X\si'=0\}\subset T_m(J^{-1}(\mu))\ll{chardist}}
provided the dimension of $K_m$ remains constant for all $m\in
J^{-1}(\mu)$. It follows that $K$ is integrable and if
$M=J^{-1}(\mu)/K$ is a manifold then $\si'$ projects onto a well defined
symplectic structure \si on $M$. This new symplectic space $(M,\si)$
is called the {\em reduction} of $(J^{-1}(\mu),\si')$.

The link between the symplectic space $M$ and the symplectic leaves
$J^{-1}(\mu)/G_\mu$ found earlier, is that, in general, $K=(G_\mu)_0$,
the identity component of $G_\mu$. When $G$ is compact or semisimple,
the isotropy group $G_\mu$ is connected \ll{Hconnected}\ci{GS} so
$K=G_\mu$ in agreement with our earlier approach. For the special case
$S=T^*G$, each $\be\in T^{*}_{x}G$ can be identified with the
one-form $\la^{*}_{x}\be\in T^{*}_{e}G\iso\gs$ where $\la^{*}_{x}$ is
the pull back of the left action $\la_{x}y=xy$. This gives us the
[left] parallelization
\bea
\TS{G} & \to & G\x\gs\nn\\
\be &\to & (x,\la^{*}_{x}\be)_{_L}\ll{lefttriv}.
\eea
Let $\{d^a\}$, $a=1,\ldots,d_G={\rm dim\ }G$, be a basis of \gs and
define $\th^a(x)=\la^*_{x^{-1}}d^a$ so $\{\th^a(x)\}$ form a basis for
the left invariant one-forms on $G$. Any element $\be\in T^*_gG$ can
be expanded as $\be=p_a\th^a(x)$ and in the above parallelization this
corresponds to $\be\to (x,p_a d^a)_L$. Hence we can use the $\{p_a\}$ as
coordinates on $T^*_gG$ which are globally valid. For future reference,
the canonical one-form in this coordinate system is
\eq{\th_0(x,p)=p_a\th^a(x).\ll{Liouville}}
(Note  that similarly there is a right parallelization of $T^*G\iso
G\x\gs$ via $\be \to(x,\rh^*_x\be)_R$ where \rh denotes the right
action of $G$ on $G$.)

If $G$ acts on the right of $S=T^*G$ then it follows from (\ref{genmommap})
that an equivariant momentum map is given by $J_R:T^*G \to \gsm$ with
$J_R(x,p)_{_L}= p$. Clearly, $J^{-1}(\mu)=G\x\{\mu\}\iso G$. Thus $G$
can be regarded as a
\ll{presymplectic}
presymplectic manifold with $K_x=\{L_V(x):V\in\g_\mu\}$ where $\g_\mu$
denotes the Lie algebra of $G_\mu$ and $L_A$ denotes the left
invariant vector field on $G$ generated by $A\in\g$, i.e., for
$f\in\cinf{G}$,
\eq{(L_Af)(x)=\left.\frac{d}{dt}f(xe^{tA})\right|_{t=0}.}
 The reduction of $G$ gives, as expected, the
manifold $G/G_\mu\iso\calO_\mu$ when $(G_\mu)_0=G_\mu$.

\ssect{Mechanical $H$-systems}
\ll{Hsystem}
We work in the setting of what Smale \ci{Smale} calls a simple
mechanical $H$-system. This means that we have a symplectic
manifold $T^*N$, together with a right action of $H$ on $N$ ($H$ acts
on $T^*N$ by cotangent lift), a Riemannian metric on $N$ which is
$H$-invariant and a Hamiltonian, $H_0:T^*N\to\Bbb R$, of the form
\eq{H_0(n,p)=\frac{1}{2}\parallel p\parallel^2_n+V(n),\ll{genH}}
where $\parallel \cdot\parallel_n$ is the norm induced on $T^*_nN$,
and where $V$ is an $H$-invariant potential.  We assume that $H$ acts
freely on $N$ so that we can regard $N\to Q=N/H$ as a principal fibre
bundle as described in section \ref{intro}. Now $H_0$ is $H$-invariant
so Marsden-Weinstein reduction gives a reduced Hamiltonian system on
the reduced space $P_\mu$ (or alternatively on
$P_{\calO_\mu}$). Marsden
\ci{Mars} has given an explicit realisation of $P_\mu$ as a
submanifold of $T^*(N/H_\mu)$, where $H_\mu$ is the isotropy group of
$H$ defined in (\ref{isotgp}). The essential part of this realisation
is what Marsden calls the mechanical connection.

The {\em locked inertia tensor} $\Bbb I(n):\h\to\hs$ is defined at
each $n\in N$ via
\eq{\langle \Bbb I(n)X,Y\rangle =\langle \langle
\xi_n(X),\xi_n(Y)\rangle \rangle ,\ll{LIN}}
where $\xi(X)$ denotes the infinitesimal generator of the action of
$\h$ on $T^*N$. We identify $\Bbb I$ with the metric on $\h$. Let
$\Bbb FL:TN\to T^*N$ be the {\em Legendre transformation} for the
simple mechanical $H$-system (e.g., see \ci{AM}).
The {\em
mechanical connection} $\al:TN\to\h$ is defined by
\eq{\al(n,v)=\Bbb I(n)^{-1}(J(\Bbb FL(n,v))),\ll{mechconn}}
where $J:T^*N\to\hs$ is the momentum map for the action of $H$ on
$T^*N$.
As mentioned in the previous section, the momentum map $J:T^*N\to\hs$
for the [right] action of $H$ on $T^*N$ is provided by means of
(\ref{genmommap}). In our present notation,
\eq{\langle J(p_n),X\rangle =\langle p_n,\xi(X)\rangle ,\hst
X\in\hs.\ll{Hmommap}}
The term mechanical connection is used because \al defines a
connection on the principal bundle $N\to N/H$.
The key construction, at least from our point of view, is the one-form
$\al_\mu$ on $N$, defined by
\eq{\langle \al_\mu(n),v\rangle =\langle \mu,\al(n,v)\rangle\ll{oneformalmu},}
i.e., $\al_\mu=\mu\crc\al$.
This one-form is used to define what Marsden \ci{Mars} calls the {\em
shifting map}
\eqn{hor:T^*N&\to& J^{-1}(0)\\
\be&\to &\be-\al_{J(\be)}.\ll{hormap0}}
Now $\al_\mu$ lies in $J^{-1}(\mu)$, so, if we restrict the map
$hor$ to $J^{-1}(\mu)$, and quotient by $H_\mu$, we have a map
\eq{hor_\mu:(J^{-1}(\mu))/H_\mu\to J^{-1}(0)/H_\mu\ll{hormap}}
induced by $p\to p-\al_\mu$.
Let $J_\mu$ denote the momentum map for $H_\mu\subset H$ (so
$J_\mu=J\rst \h_\mu$), then $J^{-1}(0)/H_\mu$ \ll{embedding} embeds in
$J_\mu^{-1}(0)/H_\mu\iso T^*(N/H_\mu)$. Thus the map $hor_\mu$ embeds
$P_\mu$ into $T^*(N/H_\mu)$.
The two-form $d\al_\mu$ on $N$ drops to a two-form, denoted by
$\be_\mu$, on the quotient $N/H_\mu$. (This is because $\al_\mu$ is
invariant under the action of $H_\mu$ which is the isotropy
group of $\mu$.) Let \i denote the embedding of $P_\mu$ into
$T^*(N/H_\mu)$ via $hor_\mu$, then the key result \ci{AM} is
that the symplectic form on $P_\mu$ is given by
\eq{\si=\i^*\si_0\mp \i^*\pi^*\be_\mu,\ll{gen2form}}
where $\pi:T^*(N/H_\mu)\to N/H_\mu$ is the canonical projection and
$\si_0$ is the canonical symplectic form on $T^*(N/H_\mu)$. The choice
of sign depends on the action of $H$; specifically if $H$ is a left
[right] action then the plus [minus] sign is chosen. Note that neither
of the two terms on the right hand side of (\ref{gen2form}) are, in
general, symplectic. However, the sum of the two is. Locally,
on a coordinate patch $M_A\subset N/H_\mu$, we can write
$\si=d\Th_A$. Let $b_A:M_A\to N$ be a (local) section,
then
\eq{\Th_A=-\i^*\th_0\mp \i^*\pi^*b_A^*\al_\mu\ll{oneform},}
where $\th_0$ is the globally defined canonical one-form on
$T^*(N/H_\mu)$.

 Alternatively, in the Kaluza-Klein picture as generalised by Kerner
\ci{Kerner}, we could start with a metric on $Q$ and a connection
form, $\al$, on $N$. As $H$ is compact, a bi-invariant metric exists
on $H$.  The metric on $N$ is induced by the connection. To be
precise, $\al$ defines an orthogonal decomposition $T_nN=V_n\oplus
H_n$, $n\in N$, where the horizontal subspace, $H_n$, is the kernel of
$\al_n$.  The metric on $V_n\iso\h$ is the one induced from the
bi-invariant metric on $\h$, whilst the metric on $H_n$ is the
pullback of the metric on $Q$. The metric on $N$ is thus $H$-invariant
since $\rh_{h*}H_n=H_{nh}$ (which is one of the defining properties of
a connection). Note that Marsden's construction of the mechanical
connection depends heavily on the given metric on $N$; whereas in the
Kaluza-Klein picture a given connection is used to construct a metric
on $N$. It is quite straightforward to show that if one starts with
this latter case and calculates the mechanical connection then it is
merely the connection one started with.

For a particle in a Yang-Mills field, the relevance of the connection
with regard to the symplectic leaves of $(T^*N)/H$ is that, as noted
by Weinstein \ci{Wein78}, until it is chosen there is no natural
projection of $(J^{-1}(\calO_\mu))/H$ on $T^*Q$; thus the variables
conjugate to position on $Q$ are inherently intertwined with the
`internal' variables associated with $\calO_\mu$. Physically, this
means that without a connection we cannot separate the particle's
external momentum from its own internal `position' and `momentum'
which is associated with the motion on the coadjoint orbit $\calO_\mu$.

\ssect{The symplectic leaves of $(T^*N)/H$}
\ll{symleaves2}
Due to the large number of fibre bundles that appear in our
discussion, we denote the projection map of a generic bundle $C$ with
base space $X$ by $\pi_{C\to X}$. For the special case of a cotangent
bundle $T^*X\to X$, we use $\pi_X$ for the projection map.

To identify the symplectic leaves of $(T^*N)/H$ we use the result,
due to Montgomery
\ci{Mont}, that $T^*N\iso N^\#\x\hs$, where $N^\#$ denotes the
pullback of the bundle $N$ to a bundle over $T^*Q$ using the canonical
projection $\pi_Q:T^*Q\to Q$. The bundle $N^\#$ is represented
diagrammatically as
\eq{\begin{array}{ccc}N^\#&\to& N\nn\\
\downarrow&&\downarrow\nn\\
T^*Q&\stackrel{\pi_Q}{\to}&Q.\nn
\end{array}}
 Further, the momentum map for the action of $H$,
$J:N^\#\x\hs\to\hs$, is given by $J(n,\nu)=\nu$. We now briefly review
these results.

The first step is that, as noted by Guillemin and Sternberg
\ci{GSpaper2}, $N^\#$ has a natural intrinsic realisation as
$V^0\subset T^*N$, the annihilator of the vertical bundle $V\subset
TN$ ($V_n\subset T_nN$ is the vertical subspace, i.e., it is the
subspace tangent to the fibre at $n\in N$). To see this, note that we
can write $N^\#=T^*Q\x_QN=\{(n,p)\in N\x T^*Q:\pi_{N\to
Q}(n)=\pi_Q(p)\}$, where $\pi_{N^\#\to T^*Q}[n,p]_Q=p$ and the
projection $pr: N^\#\to N$ is given by $pr[n,p]_Q=n$. We can pull $p$
back to an unique element $\ka_n=\pi^*_{N\to Q}p\in T^*_nN$, which is
then an element of $V^0_n$.  This correspondence between
$T^*_{\pi_{N\to Q}(n)}Q$ and $V_n^0$ is clearly bijective.

Just as a connection form, $\al_n: T_nN\to\h$, defines a unique separation
of $T_nN$ into the vertical subspace and horizontal subspace, the dual
of the connection form $\al^*_n:\hs\to T^*_nN$ defines a unique separation
of $T^*N$ into $N^\#$ and $\hs$. Specifically, $\al$ induces an
$H$-equivariant isomorphism $\tilde\al:N^\#\x\hs\to T^*N$ by
\eq{\tilde\al(\ka_n,\nu)=\ka_n+\al^*_n\nu,\ll{TSNtriv}}
where $\ka_n\in N^\#$ and we have identified $N^\#$ with $V^0\subset
T^*N$. Recall that $H$ acts on $T^*N$ by cotangent lift; the action of
$H$ on $N^\#\x\hs$ is the one induced by $\tilde\al$ and is given by
$\rh_h(\ka_n,\nu)=(\rh^*_{h^{-1}}\ka_n,\cod{h^{-1}}\nu)$. Note that
for $[n,p]_Q\in N^\#$, $\rh_h[n,p]_Q=[nh,p]_Q$. Also
$\tilde\al$ induces a symplectic structure on $N^\#\x\hs$ from the
canonical one on $T^*N$.

The moment map $J:N^\#\x\hs\to\hs$ for the action of $H$ on $N^\#\x\hs$
can be readily computed using the momentum map for the action of $H$
on $T^*N$ given in (\ref{genmommap}). We have, for $X\in\hs$,
$\langle J(\ka_n,\nu),X\rangle =\langle \ka_n,\xi(X)\rangle +\langle
\nu,\al_n(\xi(X))\rangle $, where \xii
denotes the infinitesimal generator of the right action of $H$ on $N$.
Hence
\eq{J(\ka_n,\nu)=\nu.}
Using the results of section \ref{momentummaps}, we can, using $\al$,
immediately identify the symplectic leaves of $(T^*N)/H$ with both
$P_\mu$ and $P_{\calO_\mu}$, where $P_\mu=N^\#/H_\mu$ and
$P_{\calO_\mu}=N^\#\x_H\calO_\mu$.

\ssect{A momentum map on the symplectic leaf $P_{\calO_\mu}$ -
identification of classical observables to be quantized}
\ll{observqz}
Recall in section \ref{intro} that we motivated the approach of
finding a group \calG which acts on the reduced phase space together with a
corresponding momentum map. This allows a subclass of observables to
be selected which we expect to quantize provided that we can find a
suitable representation of $\calG$. Isham \ci{Isham} has considered
the special case where $N$ is a Lie group $G$ (with $H\subset G$) so
that $Q=G/H$ is homogeneous. In particular he considered an action of
$G$ on the symplectic leaf $T^*(G/H)\subset(T^*G)/H$ with a
corresponding momentum map.  We would like to generalise this approach
for the present case where $G$ is replaced by the general principal
fibre bundle $N$ and we consider any symplectic leaf in $(T^*N)/H$.
The guiding principle is that the [left] action of $G$ on the bundle
$G\to G/H=Q$ commutes with the right action of $H$ on $G$. Hence this
action of $G$ determines a subgroup of the group of {\em
automorphisms} of $G$. For the general bundle $N$ this group is
denoted by Aut $N$ and consists of all diffeomorphisms, $\ph$, of $N$
which satisfy, for all $h\in H$,
\eq{\ph(n)h=\ph(nh).}
Note that such a \ph determines a diffeomorphism of $Q$, $\bar\ph\in
{\rm Diff\ }Q$,
via
\eq{\bar\ph(\pi(n))=\pi(\ph(n))\ll{diffeodef},}
where $\pi:N\to Q$ is the bundle projection. In the general case there
is no natural finite-dimensional subgroup of Aut $N$, thus we are forced
to consider the whole group.

We regard the Lie algebra of Diff~$N$ to be the set of all complete
vector fields on $N$. Unfortunately the commutator of two vector
fields $[A_1,A_2]=-[A_1,A_2]_{LB}$, where the subscript LB denotes the
Lie bracket of the two elements of the Lie algebra. Thus, in order to
differentiate between the two brackets we will continue to use this
subscript when considering Diff~$N$ (and Aut~$N$).

Drawing on Guillemin and Sternberg's treatment \ci{GS} of the action
of the semi-direct product group Diff~$N\lx\cinf{N}$ on $T^*N$ we
consider the subgroup Aut~$N\lx\cinf{Q}\subset {\rm Diff\ }N\lx\cinf{N}$.
The group law on Aut $N\lx\cinf{Q}$ being
$(\ph_1,f_1)\cdot(\ph_2,f_2)=(\ph_1\crc\ph_2,f_1+f_2\crc\bar\ph_1^{-1})$
and  the Lie algebra is
\eq{[(A_1,f_1),(A_2,f_2)]_{LB}=([A_1,A_2]_{LB},-A_1f_2+A_2f_1).}
Here we have identified $\calL(\cinf{Q})$ with $\cinf{Q}$. Now
Aut~$N\lx\cinf{Q}$ acts symplectically on $T^*N$ (because
Diff~$N\lx\cinf{N}$ does) and the action is given by
\eq{\ta_{(\ph,f)}\be_n=\ph^{-1*}(\be_n)-(d\pi^*f)_{\ph(n)},\hst\be_n\in
T^*_nN.\ll{taction}}
This choice of the group ${\rm Aut\ }N\lx\cinf
Q$, whose action on $T^*N$ we are interested in, agrees with that of
Landsman \ci{LandSD}.

Guillemin and Sternberg \ci{GS} give the equivariant momentum map,
$J:T^*N\to\calL({\rm Aut\ }N\lx\cinf{Q})^*$, for this action as
\eq{\langle J(p_n),(A,f)\rangle =\langle p_n,A\rangle
+\,\pi^*f(n).\ll{THEmommap}}
Now suppose $J(p'_{n'})=J(p_n)$. Clearly $\pi(n')=\pi(n)$ and hence
$n'=nh$ for some $h\in H$. We thus have
$\langle p'_{nh},A_{nh}\rangle =\langle p_n,A_n\rangle.$
But $\calL({\rm Aut\ }N)$ consists of all smooth vector fields on $N$
that are $H$-invariant (e.g., see \ci{GS}), i.e., they satisfy
$\rh_{h*}(A_n)=A_{nh}$,
for the flow of such a vector field consists of transformations
belonging to Aut~$N$. Hence
$\rh_h^*p'_{nh}=p_n$.
Thus the fibres of $J$ are generated by the right action of $H$.
Further the action $\ta$ defined in (\ref{taction}) commutes with the
right action of $H$; this explains the reasoning behind choosing
$\pi^*\cinf{Q}$ rather than $\cinf{N}$. Thus $\ta$ drops to an action
$\bar\ta$ on $(T^*N)/H$.

In passing, we note that there is a relation between the momentum map
$J$ and the dual pairs of section \ref{dualpairs}. Specifically, let
$J_R$ be the momentum map for the right action of $H$ as given in
(\ref{Hmommap}). We know from section \ref{momentummaps} that we have
the dual pair
\eq{\hs\stackrel{J_R}{\gets}T^*N\stackrel{\pi}{\to}(T^*N)/H.}
Now note that $J(T^*N)$ is finite-dimensional. Further, using
(\ref{THEmommap}), we can identify $J(T^*N)$ with $M=\{(\be,q)\in
T^*N\x Q:\pi_{T^*N\to Q}(\be)=q\}$ where $q\in Q$ is regarded as an
element of $\calL(\cinf Q)^*$ via $\langle q,f\rangle=f(q)$ for
$f\in\calL(\cinf Q)\iso\cinf Q$.
The elements of the space
$\pi^*\cinf{(T^*N)/H}$ of functions on $T^*N$ are constant on the
fibres of $J$ and hence this space coincides with the space
$J^*\cinf{M}$. Thus, we have the full dual pair
\eq{\hs\stackrel{J_R}{\gets}T^*N\stackrel{J}{\to}J(T^*N)\subset\calL({\rm
Aut\ }N\lx\cinf{Q})^*.}
We then note that $J$ induces a symplectic diffeomorphism, $\bar
J_\mu$, which maps the symplectic leaf
$P_{\calO_\mu}=(J_R^{-1}(\calO_\mu))/H\subset (T^*N)/H$ to a symplectic
leaf in $J(T^*N)\subset\calL({\rm Aut\ }N\lx\cinf{Q})^*$. Furthermore,
the map $\bar J_\mu$ is a momentum map for the action $\bar\ta$ on $(T^*N)/H$.

We have thus achieved our goal and we can now write down the classical
observables we expect to be able to quantize. These are given by
$\{\hat J_\mu(A,f):A\in \calL({\rm Aut\ }N),\ f\in\cinf{Q}\}$ where $\hat
J_\mu:\calL({\rm Aut\ }N\lx\cinf{Q})\to\cinf{P_{\calO_\mu}}$ is
given by $(\hat J_\mu(A,f))[p_n]=\langle \bar J_\mu[p_n],(A,f)\rangle $ with
$[p_n]\in P_{\calO_\mu}$, cf.\ (\ref{Jhatdef}). Recalling that
$P_{\calO_\mu}=N^\#\x_H\calO_\mu$, we then have for $p_n=(\be_n,\nu)\in
N^\#\x\calO_\mu$, using (\ref{TSNtriv})
\eq{\langle \bar J_\mu[p_n],(A,f)\rangle =\langle
\be_n+\al^*_n\nu,A\rangle +\,\pi^*f(n).\ll{Jlocation}}
Let $s$ be a local section of the bundle $N\to Q$. This allows us to
choose a specific element in each of the equivalence classes
$N^\#\x_H\calO_\mu$, so that
\eq{(\hat J_\mu(A,f))[\be_{s(q)},\nu]_H=\langle \be_{s(q)},A\rangle
+\langle \nu,\al_{s(q)}(A)\rangle +f(q).\ll{globalobserv}}
This expression simplifies if we use local coordinates.  Now
$N^\#/H=T^*Q$, and locally $P_{\calO_\mu}$ is like
$(N^\#/H)\x\calO_\mu$. Thus, let
$(h^1,\ldots,h^{d_H},q^{d_H+1},\ldots,q^{d_N})$ be local coordinates
on $N$, where $(q^{d_H+1},\ldots,q^{d_N})$ are coordinates on $Q$ and
$(h^1,\ldots,h^{d_H})$ are coordinates on the fibre $H$. Let
$p_{d_H+1},\ldots,p_{d_N}$ be the corresponding components of
covectors on $T^*Q$. Then we can label a point $[\be_{s(q)},\nu]_H$ in
$N^\#\x_H\calO_\mu$ by
$(q^{d_H+1},\ldots,q^{d_N},p_{d_H+1},\ldots,p_{d_N},\nu)$. So we can
write $\al_{s(q)}(A)=X(q^{d_H+1},\ldots,q^{d_N})$ where
$X\in\h$, together with $(\pi_{N\to
Q*}A)_q=v^\ga(q^{d_H+1},\ldots,q^{d_N})\cdd{}{q^\ga}$ where
$\ga=d_H+1,\ldots,d_N$. Thus, setting $\be_{s(q)}=\pi^*_{N\to Q}p$
with $p=p_\ga dq^\ga$, we have
\eqn{\hat
J_\mu(A,f)[\be_{s(q)},\nu]_H&=&v^\ga(q^{d_H+1},\ldots,q^{d_N})p_\ga\,+\langle
\nu,X(q^{d_H+1},\ldots,q^{d_N})\rangle\nn\\
&&\mbox{} +\,f(q^{d_H+1},\ldots,q^{d_N}).\ll{genobservables}}
This gives the classical observables which we expect to quantize.

{\emfl{The reduced Hamiltonian}}\nopagebreak
The Hamiltonian $H_0$ on $T^*N$ drops to a reduced Hamiltonian on the
symplectic leaves of $(T^*N)/H$. In particular, when a symplectic leaf
is identified with $P_{\calO_\mu}$, the reduced Hamiltonian
$H_{\calO_\mu}$ is given by \ci{Mars}
\eq{H_{\calO_\mu}(q,p,\nu)=\frac{1}{2}\parallel p \parallel^2
+\frac{1}{2}\langle \nu,\Bbb I(s(q))\nu\rangle+V(s(q)),}
where $(q,p,\nu)$ labels locally a point in $N^\#\x_H\calO_\mu$ as
above. Denote by $A_\ga$ the element of $\calL({\rm Aut\ }N)$ such
that $\pi_{N\to Q*}A_\ga=\cdd{}{q^\ga}$ and $\al(A_\ga)=0$.
Then $\hat J_\mu(A_\ga,0)[\be_{s(q)},\nu]_H=p_\ga$. Similarly denote by
$A_I$ the element of $\calL({\rm Aut\ }N)$ such that $\pi_{N\to
Q*}A_I=0$ and $\al(A_I)=T_I$ where
$\{T_J:J=1,\ldots,d_H\}$ is a basis for $\h$. Hence $\hat
J_\mu(A_I,0)[\be_{s(q)},\nu]_H=\langle\nu,T_I\rangle$. We can then
write the reduced Hamiltonian as
\eq{H_{\calO_\mu}=\frac{1}{2}{\sf g}^{\al\be}\hat J_\mu(A_\al,0)\hat
J_\mu(A_\be,0)+\frac{1}{2}\Bbb I^{IJ}\hat J_\mu(A_I,0)\hat
J_\mu(A_J,0)+ \hat J_\mu(0,V_0).\ll{Ham3}} Here $V_0\in\cinf Q$ is
such that $\pi^*_{N\to Q}V_0=V$ while $\{{\sf g}_{\al\be}\}$ and $\{\Bbb
I_{IJ}\}$ are the metrics on $Q$ and \h respectively (${\sf
g}_{\al\be}{\sf g}^{\be\ga}=\de_\al^\ga$, $\Bbb I^{IK}\Bbb I_{KJ}=\de^I_J$).

\section{Geometric quantization}
\ll{gq}
\ll{gqreview}

We give a brief outline of the main procedures of geometric
quantization. The reader is referred to Woodhouse \ci{Wood},
Sniatycki \ci{Snki} or Puta \ci{Puta} for comprehensive expositions.

\ssect{Prequantization and Polarizations}
\ll{prequantum}
Prequantization is the process of finding the Hilbert space \calH
described in section \ref{intro} together with the map $f\to \hat f$
which links classical observables with their counterpart quantum
operators. A complex Hermitian line bundle $B$ over the symplectic
space $M$ is introduced along with a connection, $\del$, on $B$ with
curvature $\hbm\si$, where \si is the symplectic form on $M$. The
bundle $B$ is called the {\it prequantum line bundle}. An inner
product $\langle \ ,\ \rangle $ on $\Ga(B)$ (the sections of $B$) is
given by
\eq{\langle s_1,s_2\rangle =\int_M(s_1,s_2)\si^n,}
where dim~$M=2n$. We restrict \calH to be the space of
 square-integrable sections.  Each observable, $f\in\cinf{M}$,
 corresponds to the operator $\hat f$, where
\eq{\hat f s=-i\hbar\del_{\xi_f}s+fs\ll{fhataction},}
and $\xi_f$ is the Hamiltonian vector field associated to $f$. This
ensures that conditions (Qi) to (Qiii) of section \ref{intro}
hold. If $\xi_f$ is complete then $\hat f$ is essentially self-adjoint
(on a suitable domain).

Associated to each observable $f$ is a vector field $V_f$ on $B$
characterised by
\eq{\left\{\begin{array}{rclcl}\pi_{B\to P_\mu*}V_f&=&\xi_f;\\\\
\hbar\langle \tilde\Th, V_f\rangle &=&\hbar\langle \bar{\tilde\Th},
V_f\rangle &=&-f\crc\pi_{B\to P_\mu}
.\end{array}\right.\ll{observcond}}
Here $\tilde\Th$ is the
connection one-form on the prequantum bundle $B$ and $\bar{\tilde\Th}$
is its complex conjugate.
Let $\rh_t$ denote the flow of $\xi_f$, and $\de_t$
the flow of $V_f$. For a section $s\in\Ga(B)$ a linear ``pullback''
action $\hat\rh_t:\Ga(B)\to\Ga(B)$ can be defined by
\eq{\de_t(\hat\rh_ts(m))=s(\rh_t(m))\ll{flowsdef}.}
Then, $\hat\rh_t$ is related to the quantum operator $\hat f$
 via
\eq{\left.\frac{d\hat\rh_t}{dt}\right|_{t=0}=i\hbm\hat f\ll{qopdef}.}

For a given symplectic manifold, a prequantum bundle does not always
exist. This leads to what are called quantization or integrality
conditions which determine if and when a prequantum bundle exists.
Such conditions are usually formulated as a requirement on an integral
of the symplectic form or in terms of de Rham cohomology classes.

The next step in geometric quantization, once the prequantum bundle
has been found, is to construct a {\em polarization} of the symplectic
manifold. Then the Hilbert space $\calH$ is replaced by sections which
are parallel along the polarization. Such sections are called
polarized sections. The class of observables that can be quantized is
then restricted to those for which the flow of the corresponding
Hamiltonian vector field preserves the polarization.

\ssect{Quantization on coadjoint orbits}
\ll{Hqz}
Much has been written on the subject of geometric quantization on
coadjoint orbits; e.g., see Woodhouse \ci{Wood}, and Baston and
Eastwood \ci{BE}. We briefly outline the main steps.

Let $H$ be a compact connected Lie group with Lie algebra $\h$; let
$\mu\in\hs$ and let $\calO_\mu\subset \hs$ denote the coadjoint orbit
of $\mu$. Recall, that in section \ref{momentummaps},
p.~\pageref{presymplectic}, we saw that we could regard $H$ as a
presymplectic manifold. Also, we noted that for compact $H$ (assumed
here), $H_\mu$, the isotropy group of $\mu$, is connected; then the
reduction of $H$ by the left action of $H_\mu$ gives $H_\mu\bs
H\iso\calO_\mu\subset\hs$.  The symplectic 2-form, $\om_\calO$, on
$H_\mu\bs H$ is given by $\pi^*\om_\calO=\om_\mu$ where $\pi:H\to
H_\mu\bs H$ is the projection and $\om_\mu$ is the restriction of the
canonical 2-form on $T^*H$ to $H\x\{\mu\}\iso H$ where we are working
with the right trivialization of $T^*H$. Thus, from (\ref{Liouville}),
$\om_\mu=-d\th_\mu$ where $\th_\mu(h)=\rh^*_{h^{-1}}\mu$.

Having detailed the symplectic manifold $(H_\mu\bs H, \om_\calO)$, the
next step is to construct the prequantum bundle. Drawing on Woodhouse
\ci{Wood}, Kostant's formulation of the integrality condition
on $\om_\calO$ can be expressed as the requirement that $i\hbm\mu$
should be the gradient at $e$ of a homomorphism $\ch_\mu:H_\mu\to\Bbb T$,
where $\Bbb T$ is the circle group.

The prequantum line bundle, $B$, is given by $B=H\x_{H_\mu}\Bbb C$,
i.e., $H\x\Bbb C$ modulo the equivalence relation $(h,z)\sim(h_\mu
h,\ch_\mu(h_\mu)z)$ for $h\in H$, $h_\mu\in H_\mu$. This bundle has a
connection whose curvature is $\hbm\om_\calO$. The connection can be
either considered in the light of \ci{Wood} or in the following
manner. The principal bundle $H\to H_\mu\bs H$ has the canonical
$H$-invariant (under right action) connection (e.g., see \ci{KNI}); by
the assumption on the integrality of $\mu$, there is a representation
of $H_\mu$ into U(1). Under the derivative of this representation, the
canonical connection becomes a connection on $B'$ with curvature
$\hbm\om_\calO$.

We can identify
the sections of $B$ with functions $\ph:H\to \Bbb C$ satisfying
\eq{\ph(h_\mu h)=\ch_\mu(h_\mu)\ph(h).}
There is an induced representation, $\pi_\mu$ of $H$, on these functions
defined by
\eq{(\pi_\mu(h')\ph)(h)=\ph(hh').}
An inner product is given by
\eq{\langle \ph_1,\ph_2\rangle =\int_{H_\mu\bs
H}d([h]_{H_\mu}) \langle \ph_1(h),\ph_2(h)\rangle _{\Bbb C}.}

We only require two more facts concerning the quantization on
coadjoint orbits, namely that there is a positive $H$-invariant
polarization on $H_\mu\bs H$ and that the representation $\pi_\mu$
acting on the polarized sections of $B$ is irreducible. The first of
these is a standard result \ci{Wood} whilst the second, however, is
only given (refs.\ \ci{Wood} and \ci{BE}) when it is assumed that $H$
is simply connected. The extension to the case where this assumption
is no longer made is just the application of a series of standard
results. The main one being that the Borel-Weil theorem \ci{Knapp}
holds without this assumption. We now briefly give the details.

Closely following \ci{Wood}, except where indicated, let
$T$ denote a maximal torus in $H$. There is an arbitrariness in
choosing $T$ and we may use this freedom to ensure that $T\subset
H_\mu$. Let $\De$ denote the set of roots of \h and, for $\al\in\De$,
let $\g_\al$ denote the corresponding eigenspace. For $A\in\g_\al$, $A\neq0$,
define $Z_\al=\frac{1}{2}i[\bar A,A]$, rescale $A$
such that $\al(Z_\al)=i$ and let $\De^+_\mu$ be the subset of $\De$
such that $\langle \mu,Z_\al\rangle >0$. Now set
$\bb=\bt_{\Bbb C}\oplus_{\al\not\in\De^+_\mu} \g_\al$. We can define a
complex distribution $P'$ on $H$ by $P'_h=\rh_{h*}\bb$. The projection
of $P'$ onto $H_\mu\bs H$ is a positive K\"ahler polarization,
$P^\calO$, of $H_\mu\bs H$ \ci{Wood}. Further, this polarization is
$H$-invariant, which means that for each $[s]\in H_\mu\bs H$ we have
$\rh_{h*}P^\calO_{[s]}=P^\calO_{[s]h}$, where $H$ acts naturally on
the right of $H_\mu\bs H$.

We now return to the representation, $\pi_\mu$ of $H$, mentioned
above. Firstly the integrality condition on \muu means that $\langle
\mu,A\rangle $ is an integer multiple of $\hbar$ for every $A\in\h$
such that $e^{2\pi A}$ is the identity \ci{Wood}. This means that, in
the terminology of weight theory, \muu is {\em analytically
integrable} \ci{Knapp}. Also, since $\langle
\mu,Z_\al\rangle >0$ for all $\al\in\De^+_\mu$, \muu is said to be
{\em dominant} \ci{Knapp}. Now, as a complex manifold $H_\mu\bs H$ is
the same as the homogeneous space $\calB\bs H_{\Bbb C}$, where $\calB\subset
H_{\Bbb C}$ is the subgroup generated by
\bb and $H_{\Bbb C}$ is the complexification of $H$. Further, the
representation $\ch_\mu$ of $H_\mu$ extends to \calB \ci{Wallach}, so
that we can identify the {\em polarized} sections of $B$ with
functions $\ph:H_{\Bbb C}\to\Bbb
C$ satisfying:
\begin{description}
\item[(i)] \ph is holomorphic;
\item[(ii)] $\ph(bx)=\ch_\mu(b)\ph(x)\ \ \fa b\in \calB,\ x\in H_{\Bbb C}$.
\end{description}
The representation $\pi_\mu$ acts in the same manner as before on
these polarized sections and by the Borel-Weil theorem \ci{Knapp},
$\pi_\mu$ is an irreducible unitary representation of $H$. Further, by
choosing an appropriate value of $\mu$, all finite irreducible unitary
representations of $H$ are obtained in this way. The reader is
referred to \ci{Wood} for a discussion on the value of \muu which
generates a given representation.

\sect{Geometric quantization of the symplectic leaves of $(T^*N)/H$}
\ll{gqgen}
Recall that the reduced phase space of our constrained mechanical
system can be identified with a symplectic leaf of $(T^*N)/H$. We now
apply the technique of geometric quantization to these symplectic spaces.
\ssect{The prequantum line bundle $B\to P_\mu$}
\ll{preqbundle}
We require a (complex) hermitian line bundle $B\to P_\mu$ and a
connection $\del$ on $B$ with curvature $\hbm\si$, with $\si$ given in
(\ref{gen2form}). In particular, we saw in section \ref{Hsystem} that
the symplectic form on $P_\mu$ was built from the 2-form $\be_\mu$
defined on $N/H_\mu$ and the canonical 2-form on $T^*(N/H_\mu)$. Thus,
we aim to find a line bundle, $B'$, over $N/H_\mu$ and a connection on
$B'$ with curvature $\hbm\be_\mu$. We can pullback $B'$ by $\pi$ to
form $\pi^*B'\to T^*(N/H_\mu)$, where $\pi$ is the same as in
(\ref{gen2form}). The tensor product bundle formed from $\pi^*B'$ and
the trivial bundle $B_0=T^*(N/H_\mu)\x\Bbb C$ will yield a line
bundle, $B_1=\pi^*B'\otimes B_0$, with curvature the sum of the
curvatures of $\pi^*B'$ and $B_0$. (The simple expression for the
curvature is a consequence of the additivity of the Chern character
under the formation of tensor product bundles.) Now $B_0$ admits a
connection with curvature $\hbm\si_0$, thus, by considering
(\ref{gen2form}), $B=\i^*B_1$ will be a line bundle over $P_\mu$ with
the desired connection, where $\i$ is defined just before (\ref{gen2form}).

The key point in constructing the line bundle $B'\to N/H_\mu$ is that
$\al$ defines a connection, $\al'$, on $N\to N/H_\mu$ via
$\al'=pr\crc\al$ where $pr:\h\to\h_\mu$ is the projection relative to
the metric on $H$. A representation, $\ch_\mu$, of $H_\mu$ into U(1)
then allows us to define the associated line bundle
$B'=N\x_{H_\mu}\Bbb C$ (where $(n,z)\sim(nh_\mu,\ch_\mu(h_\mu^{-1})z)$
for $h_\mu\in H_\mu$) with a corresponding connection. Of course
there is a restriction on $\ch_\mu$ if $B'$ is to have the desired
connection. Interestingly, the condition on $\ch_\mu$ is the same as
Kostant's formulation of the integrality condition for the
quantization of coadjoint orbits described in section \ref{Hqz}, i.e.,
$i\hbm\mu$ should be the gradient at $e$ of a homomorphism
$\ch_\mu:H_\mu\to \Bbb T$ where $\Bbb T$ is the circle group. To see
this, note that  $\ch_\mu$ defines a representation
of $H_\mu$ into U(1) and its derivative defines a representation
$\ch'_\mu:\h_\mu\to\Bbb C$ which is given by $\ch'_\mu(A)=i\hbm\langle
\mu,A\rangle $.  Under this derivative the connection $\al'$ gives a
connection on the associated bundle $B'$ with curvature $\hbm\be_\mu$,
where $\be_\mu$ denotes the two-from $d\al_\mu$ dropped to
$N/H_\mu$. Specifically, let $\calA$ be the local expression for
$\al'$, then from the definition of a covariant derivative
\eq{\del_X(\ps s)=(X(\ps)+\ch'_\mu(\calA(X))\ps)s,\hst
X\in\Ga(T(N/H_\mu)).}  Here, $s$ denotes the unit section and $\ps$ is
a complex valued function. But $\ch'_\mu(\calA(X))=i\hbm\langle
\calA_\mu,X\rangle $ where $\calA_\mu=\langle \mu,\calA\rangle
$. Hence
\eq{\del_X(\ps s)=(X(\ps)+i\hbm\langle \calA_\mu,X\rangle \ps)s,}
so $\calA_\mu$ determines a connection with curvature $\hbm
d\calA_\mu=\hbm\be_\mu$. For future reference we note that we can
identify sections of $B'\to N/H_\mu$ with functions $\ga:N\to \Bbb C$
such that
\eq{\ga(nh_\mu)=\ch_\mu(h_\mu^{-1})\ga(n)\hst\fa h_\mu\in H_\mu.\ll{preqcond}}

For completeness we relate our approach to that of Woodhouse
\ci[proposition 8.4.9]{Wood} for the construction of the prequantum
line bundle for the reduction of a symplectic manifold.
Specifically, in our present notation, Woodhouse defines the line bundle to
be $N\x\Bbb C$ quotiented by the equivalence relation
$(n_1,z_1)\sim(n_2,z_2)$ if $\pi(n_1)=\pi(n_2)$ and $z_2=z_1\exp
(i\hbm\int^{n_2}_{n_1}\al_\mu)$. Here $\pi:N\to N/H_\mu$ is the
projection map and the precise path of the integral does not matter
since it supposed that $\al_\mu$ satisfies the integrality condition
$\frac{1}{2\pi\hbar}\int_\ga\al_\mu\in\Bbb Z$ whenever \ga is a closed
curve in a fibre of $N\to N/H_\mu$. (Note that Woodhouse's construction does
not require the 2-form $d\al_\mu$ to be symplectic.)

{}From the defining properties of a connection, it immediately follows
that, for $A\in\h_\mu$, $\langle \al_\mu,\xi(A)\rangle =\langle
\mu,A\rangle $ and, additionally using the $H_\mu$ invariance of
$\mu$, $\rh^*_{h_\mu}\al_\mu=\al_\mu$, where $h_\mu\in
H_\mu$. Considering the equivalence relation defined above, clearly
$n_2=n_1h_\mu$ for some $h_\mu\in H_\mu$. Thus, for $h_\mu=e^A$, where
$A\in\h_\mu$, we have
\eq{\exp\left(
\frac{i}{\hbar}\int^{n_1e^A}_{n_1}\al_\mu\right)
=\exp\left(\frac{i}{\hbar}\langle
\mu,A\rangle \right).\ll{chidef0}}
Recall that $H_\mu$ is connected (see section \ref{symleaves}, p.\
\pageref{Hconnected});
hence we can define $\ch_\mu:H_\mu\to \Bbb C$ by
\eq{\ch_\mu(h_\mu)=\exp\left(
\frac{i}{\hbar}\int^{nh_\mu}_{n}\al_\mu\right).\ll{chidef2}}

Note that the right hand side of (\ref{chidef2}) is independent of $n$
and so $\ch_\mu$ is well defined. Thus, the  integrality condition is
equivalent to $\ch_\mu$ being a single valued function on $H_\mu$.
Further,
\eq{\ch_\mu(h_\mu e^{tA})=\ch_\mu(h_\mu)\ch_\mu(e^{tA}),}
and hence that $\ch_\mu(h_\mu h'_\mu)=\ch_\mu(h_\mu)\ch_\mu(h'_\mu)$
for all $h'_\mu,h_\mu\in H_\mu$. Now, by noting that
\mbox{$\langle d\ch_\mu,\xi(A)\rangle $}$=i\hbm\ch_\mu(h_\mu)\langle
\mu,A\rangle $, we see that $\ch_\mu$
is a homomorphism $\ch_\mu:H_\mu\to\Bbb T$ whose gradient at $e$ is
$i\hbm\mu$. Reversing the argument, it can be seen that the converse
holds. Thus we find the same condition on $\ch_\mu$ as before.

Note that, given $\al$, our construction defines a unique line bundle
$(B',\del)$. This is in contrast with the usual situation in geometric
quantization because there the symplectic 2-form, $\si$, is the
starting point and this does not define a unique one-form \th such
that $d\th=\si$; the construction of the line bundle uses the 1-form
$\th$, and thus this process does not, in general, give a unique bundle
unless the symplectic space is simply connected (e.g., see \ci{Wood}).
Our approach avoids this problem because we start with $\al_\mu$
rather than $\be_\mu$.

Recalling the comments made in the opening paragraph of this section
we have now proved
\begin{theorem}
Let $B_0$ be the trivial bundle $T^*(N/H_\mu)\x\Bbb C$ with a
connection determined by the local connection form $-\hbm\th_0$, where
$\th_0$ is the globally defined canonical one-form on
$T^*(N/H_\mu)$. Then the prequantum line bundle $B\to P_\mu$ is given
by $\i^*(\pi^*B'\otimes B_0)$ where $\pi:T^*(N/H_\mu)\to N/H_\mu$ is
the canonical projection and $B'=N\x_{H_\mu}\Bbb C$ is the line bundle
given above.
\end{theorem}
For clarity and for future reference we note that the bundle $B$ has
local connection one-forms $-i\hbm(\i^*\th_0+\i^*\pi^*\ga^*\al_\mu)$ on
$\pi^{-1}(M)$, $M\subset N/H_\mu$, where $\ga:M\to N$ is a local
section.

If we use principal bundles rather than their associated vector
bundles, the trivialization $T^*N\iso N^\#\x\hs$ allows more explicit
forms for the various bundles just described to be given.
The key point is to realise that the map $hor$ defined
in (\ref{hormap0}) is no more than the projection $T^*N\iso N^\#\x\hs\to
N^\#$. Also, by considering $\pi^*_{N/H_\mu}N$ to be the annihilator
of the vertical bundle of $N\to N/H_\mu$, we see that
$\pi^*_{N/H_\mu}N\iso N^\#\x\ns$, where $\ns\subset\hs$ is
defined to be the annihilator of $\h_\mu\subset\h$. Then, the pullback
bundles of $N\to N/H_\mu$ are given by the following diagram.
\eq{\begin{array}{ccccc}
i^*\pi^*_{N/H_\mu}N\iso N^\#&\to&\pi^*_{N/H_\mu}N\iso
N^\#\x\ns&\to&N\\\nn\\
\downarrow&&\downarrow&&\downarrow\\\nn\\
P_\mu\iso N^\#/H_\mu&\stackrel{\i}{\to}& T^*(N/H_\mu)\iso
N^\#\x_{H_\mu}\ns&\stackrel{\pi_{N/H_\mu}}{\to}&N/H_\mu
\end{array}}
The bundle $\i^*\pi^*_{N/H_\mu}B'$ is given by $N^\#\x_{H_\mu}\Bbb C$.
(The prequantum bundle $B$ has the same structure but the connection
is not the one induced from $B'$.)

{\emfl{The Aharanov-Bohm effect}}\nopagebreak
Briefly \ci{AB,Wood,Nak}, Aharanov and Bohm considered the case of a
particle with charge $e$ moving in the region outside an [infinitely]
long cylinder, so that the configuration space, $Q$, of the system is
no longer simply connected. Inside the cylinder there is a
non-vanishing magnetic field; even if the magnetic field in $Q$
vanishes, there is no gauge in which the magnetic vector potential, $\calA$,
vanishes in $Q$. It is found that the potential influences the motion
of the particle, in that the phase change of the wave function of the
particle around a closed loop surrounding the cylinder is not zero,
but is given by
\eq{\exp\left(\frac{-ie}{\hbar}\oint \calA_a dq^a\right).\ll{phase}}
The Aharanov-Bohm effect in the context of geometric quantization is
well understood \ci{HorvAB1,HorvAB2}. We now quickly show how our
approach reproduces the expected results.

The phenomena of electromagnetic fields is described by a U(1) gauge
theory. Thus we have $H={\rm U}(1)$ and hence $H_\mu={\rm U}(1)$
also. A magnetic vector potential $\calA$ corresponds to a connection
\al on the bundle $N\to Q=N/H$ (i.e., \calA is the local form on $Q$
for $\al$). Let $\ga:[0,1]\to N/H_\mu=N/H$ be a closed loop. Denote by
$\tilde\ga$ the horizontal lift of $\ga$ to $N$ with respect to the
connection $\al'$. (Note that $\al'=\al$ here since $H_\mu=H$.) Define
$h_\mu\in H_\mu$ by
\eq{\tilde \ga(1)=\tilde\ga(0)h_\mu,}
then we have, by a direct consequence of the construction of
$\tilde\ga$,
\eq{h_\mu=\exp\left(-\oint\calA_adq^a\right)\ll{hdef}.}
Now the phase change in $\ps$, a section of the bundle $B'$, on going
round the loop \ga by parallel transport is just $\ch_\mu(h_\mu)$.
{}From (\ref{chidef2}) and (\ref{chidef0}), we have immediately
\eq{\ch_\mu(h_\mu)=\exp\left(\frac{-i\mu}{\hbar}\oint\calA_adq^a\right),}
where we have identified $\calL({\rm U}(1))$ with $\Bbb R$. This agrees with
(\ref{phase}) since $\calO_\mu=\{\mu\}\in\hs$ is identified with the
charge $e$ of the particle. Hence our construction {\em automatically}
gives the physically correct choice of the prequantum line bundle. At
this stage it is not clear how this generalises to non-Abelian gauge
groups. We will return to this as the end of section \ref{poldsections}.

\ssect{A polarization for $P_\mu$}\ll{polPmu}
Recall, that we may consider $P_\mu\iso P_{\calO_\mu}=
N^\#\x_H\calO_\mu$. Now, we saw that, in section \ref{Hqz},
the coadjoint orbit $\calO_\mu\iso H_\mu\bs H$ has a natural
$H$-invariant positive K\"ahler polarization $P^\calO$. (The
$H$-invariance of the polarization means that for each $[s]\in H_\mu\bs
H$, we have $\rh_{h*}P^\calO_{[s]}=P^\calO_{[s] h}$, where $H$ acts
naturally on the right of $H_\mu\bs H$.)

For a cotangent bundle $T^*Q\to Q$, a natural polarization is given by
the complexified vertical subspace at each $u\in T^*Q$ \ci{Wood},
i.e., the complexified subspace of $T_u(T^*Q)$ which is tangent to the
fibre. (This is called the vertical polarization.) In a similar
manner, we may define an integrable complex distribution $P_0$ (a
sub-bundle of the complexified tangent bundle) on the bundle $N^\#\to
N$ simply by taking the complexified subspace of the tangent space
which is tangent to the fibre.  The fibre of the bundle $N^\#\to N$ at
$n\in N$ corresponds to $V_n^0$, the annihilator of the vertical
subspace $V_n$. Now $\rh_{h*}V_n=V_{nh}$, so that
$\rh^*_{h^{-1}}V_n^0=V^0_{nh}$. Hence, the distribution $P_0$ is
invariant under the right action of $H$.

The direct sum of the polarization, $P^\calO$, on $\calO_\mu$ and the
distribution, $P_0$, on $N^\#$ gives a new $H$-invariant
distribution, $P'$, on $N^\#\x\calO_\mu$. The projection of $P'$ onto
$N^\#\x_H\calO_\mu$ is the distribution given by
\eq{P_m=pr_*P'_u,}
where $pr:N^\#\x\calO_\mu\to N^\#\x_H\calO_\mu$ is the projective
map and $u$ is any element of $pr^{-1}(m)$. The precise choice of $u$
is irrelevant since both $P^\calO$ and $P_0$ are invariant under the
action of $H$. For suppose $u_1,u_2\in pr^{-1}(m)$ then $u_1=u_2\cdot
h$ for some $h\in H$. By $H$-invariance,
$pr_*P'_{u_1}=pr_*\rh_{h*}P'_{u_2}=pr_*P'_{u_2}$ as required.

\begin{theorem}
The distribution $P$ is a
polarization for $P_{\calO_\mu}$.
\end{theorem}
{\em Proof.} Firstly, $P$ and $P+\bar P$ are involutory since the
push-forward map $pr_*$ preserves commutators. Further, by
construction we see that $P$ is smooth and also $P_m \cap \bar
P_m=pr_*(P_0\oplus \{0\})$. Thus $P_m \cap \bar P_m$ is of constant
dimension. We finally need to show that $P$ is maximally
isotropic. Clearly $P$ has the right dimension since ${\rm dim\ }
P={\rm dim\ } N-{\rm dim\ }\hs+1/2\ {\rm dim\ }
\calO_\mu$ whilst $ {\rm dim\ } N^\#\x_H\calO_\mu=2{\rm dim\ } N-{\rm
dim\ }\hs- {\rm dim\ }H+{\rm dim\ }\calO_\mu$. To show that $P$ is
isotropic we need the expression for the symplectic form on
$P_{\calO_\mu}$ given in (\ref{symformdiff}).  Since we are using the
trivialization $T^*N\iso N^\#\x\hs$, $\Om$ should be replaced by
$\tilde\al^*\Om$ which is the induced symplectic form on
$N^\#\x\hs\iso T^*N$. We have
\eqn{\Om_\calO(P_m,P_m)&=
&(j^*_\calO\tilde\al^*\Om)(P'_u,P'_u)-(J^*_\calO\om_\calO)(P'_u,P'_u)\\
&=&0,} where to justify
$\Om(\tilde\al_*j_{\calO*}P'_u,\tilde\al_*j_{\calO*}P'_u)=0$, we
consider local canonical coordinates $\{q^i,p_j\}$ on $T^*N$. Then, we
see that $\tilde\al_*j_{\calO*}P'_u$ is spanned by $\{\cdd{}{p_j}\}$,
and thus the desired result. Hence $P$ is maximally isotropic and is a
polarization for $P_{\calO_\mu}\iso P_\mu.\ \blacksquare$

\ssect{Quantization}
\ll{gengq}
Having found a polarization for $P_\mu$, the standard approach in
geometric quantization is to replace the pre-Hilbert space of smooth
square-integrable sections of the prequantum line bundle, $B$, with
the subspace of square-integrable polarized sections of $B$.  The
quantum operator corresponding to a classical observable is defined on
the polarized sections of the prequantum line bundle $B$.  However,
these sections are not square-integrable on $P_\mu$. Thus, in a manner
analogous to that described in \ci{Wood}, we alter the
quantization process so that we integrate over $Q$ rather than
$P_\mu$.

Briefly, let $\pi:P_\mu\to Q$ be the canonical projection and let
$\De_Q\to Q$ denote the line bundle $\La^nT^*_{\Bbb C}Q$. (Here
$\La^nV$ is the $n$-fold exterior power of a vector space $V$ and
$n={\rm dim}\ Q$.) Then, define $K_D=\pi^*\De_Q\subset \La^nT^*_{\Bbb
C}P_\mu$. The bundle $\De_Q$ is trivial so $K_D$ is too. Thus, we can
define $\de_D=\sqrt K_D$. We now replace $B$ by $B_K=B\otimes \de_D$
and consider polarized sections of this bundle. In terms of the bundle
$E$, we can view sections of the new bundle $B_K$ as sections of the
bundle $E\otimes\sqrt \De_Q\to Q$. Sections of this bundle are of the
form $\tilde s=s\et$ where $s\in\Ga(E)$ and $\et\in\Ga(\sqrt
\De_Q)$. The inner product for such sections is
\eq{(\tilde s_1,\tilde
s_2)=\int_Q\langle s_1(q),s_2(q)\rangle _{\calH_\mu}(\et_1,\et_2),\ll{innerp}}
where $(\et_1,\et_2)=\et_1\bar\et_2\in \De_Q$.

The quantum operator, $\tilde f$, corresponding to a classical
observable, $f$, is given by \ci{Wood}
\eq{\tilde f\tilde s=\hat f(s)\nu-\frac{1}{2}i\hbar s ({\rm div}\
\xi_f)\et,\ll{ftildeaction}}
where $\tilde s=s\et$ and div\ll{divdef} is defined with respect to
\et via $\tilde \calL_V\et^2=({\rm div}\ V)\et^2$. (Here, $\tilde \calL$
denotes the Lie derivative.) However, only certain observables can be
quantized. Specifically, the flow of $\xi_f$ must preserve the
polarization and, additionally, we are interested in the case when
$\xi_f$ is complete so that the operator $\tilde f$ is essentially
self-adjoint (on a suitable domain). For a vertical polarization of
$T^*Q$, the form of such an observable is
\eq{f=v^i(q)p_i+u(q)\ll{realobserv},}
where $v\in\Ga(TQ)$ and $u\in\cinf Q$ \ci{Wood}. For a K\"ahler
polarization, $\xi_f$ must be a Killing vector \ci{Wood}. We now
consider a special case of the latter, namely, a K\"ahler polarization
on a coadjoint orbit, $\calO_\mu\iso H_\mu\bs H$. We can regard $X\in
\h$ as an element of $\cinf{\calO_\mu}\subset\cinf{\hs}$ via
$X(\nu)=\langle \nu,X\rangle $ where $\nu\in\calO_\mu\subset\hs$. The
Hamiltonian vector field for such an observable is $\pi_{co}(X')$,
where $\pi_{co}$ in this context means the derived Lie algebra
representation of the coadjoint action of $H$. Clearly, such a vector
is a Killing vector since the metric on $\hs$ (induced from the one on
$\h$) is invariant under the coadjoint action of $H$.

Now the symplectic leaf $P_{\calO_\mu}=N^\#\x_H\calO_\mu$ is {\em
locally} a product of [a subset of] the cotangent bundle $T^*Q$ and
the coadjoint orbit $\calO_\mu$. Similarly, the polarization is
locally a product of a vertical polarization and a K\"ahler
polarization. Thus, using (\ref{realobserv}) and the comments above on
the observables that can be quantized for a K\"ahler polarization, we
see that, {\em crucially}, the general form of a classical observable
which can be quantized to give a self-adjoint operator agrees with
that given in (\ref{genobservables}). This agreement between the
prediction, made in section \ref{observqz} via the use of a momentum
map, of which classical observables should be quantized and the actual
observables which can be quantized via the geometric quantization
technique is striking and indeed most reassuring.

The quantum operator corresponding to the observable given in
(\ref{genobservables}) acts on sections of $B_K=B\otimes \de_D$ and
can be found using (\ref{ftildeaction}). However, there is a much more
elegant way to present the quantum operators, namely as the Lie
algebra representation derived from a representation of a Lie
group. We now explain this approach.

\ssect{Polarized sections of the prequantum line bundle}
\ll{poldsections}
In order to make the connection with induced representations we must
first represent the polarized sections of $B$ in a more transparent manner.

To begin with, consider the line bundle $B'=N\x_{H_\mu}\Bbb C\to
N/H_\mu$. Now $N/H_\mu\iso N\x_H(H_\mu\bs H)\iso N\x_H\calO_\mu$ and
we can represent sections of $B'$ by functions $\ps:N\x H\to \Bbb C$
satisfying
\eq{\left\{\begin{array}{llll}\ps(n,h')&=&\ps(nh,h'h)&\fa h\in H\nn;\\
\ps(n,h_\mu h')&=&\ch_\mu(h_\mu)\ps(n,h')&\fa h_\mu\in
H_\mu.\nn\end{array}\right. \ll{fncreln}}
{}From $\ps$ we can define a function $\ga:N\to \Bbb C$ via
\eq{\ps(n,h)=\ps(nh^{-1},e)\equiv\ga(nh^{-1}).\ll{gadef}}
It is easy to see that $\ga(nh_\mu)=\ch_\mu(h_\mu^{-1})\ga(n)$. Hence
\ga satisfies (\ref{preqcond}) and thus $\ps$ represents a section, $s$, of
$B'$. We can pull $s$ back to give a section of the prequantum line
bundle $B$. The crucial point is how to realise the polarization
condition on the $\ps$'s.

We can define a distribution on $N\x_H\calO_\mu$ in a similar manner
to that used in section \ref{polPmu}. Specifically we take the trivial (zero)
distribution on $N$ and the normal K\"ahler polarization on $H_\mu\bs
H\iso\calO_\mu$. The direct sum of the two distributions on
$N\x\calO_\mu$ projects to a distribution on $N\x_H\calO_\mu$. The key
point is that sections, $s$, of $B'$ which satisfy the
``pseudo-polarization'' condition, $\del_{\bar X}s=0\ \fa X\in
V_P(N\x_H\calO_\mu)$, pullback to {\em polarized} sections of
$B$. Further {\em all} polarized sections occur in this way. Everything
becomes clearer if we use local coordinates. Namely, if $\{q^a,p_a\}$
are local canonical coordinates for $T^*(N/H)$ and
$\{z^i\}$ are local
(complex) coordinates for $\calO_\mu$, then polarized sections of $B$
are of the form $\ph(q,z)$, i.e., holomorphic in $z$. Clearly these
correspond directly with sections of $B'$ satisfying the
pseudo-polarization condition.

In terms of the functions $\ps:N\x H\to \Bbb C$, if we set
$\ph_n(h)=\ps(n,h)$ and regard $\ph_n:H\to \Bbb C$ then the condition
that \ps will correspond to a polarized section of $B$ is that $\ph_n$
represents a polarized section of $H\x_{H_\mu}\Bbb C$ (where
$(h,z)\sim(h_\mu h,\ch_\mu(h_\mu)z)$, $h_\mu\in H_\mu$) with respect
to the K\"ahler polarization on $H_\mu\bs H$. Consequently, let $\calH_\mu$
be the completion of the pre-Hilbert space of square-integrable
polarized sections of
$H\x_{H_\mu} \Bbb C$. Note that sections of $H\x_{H_\mu}\Bbb C$ are
represented by functions $\ph:H\to
\Bbb C$ satisfying
\eq{\ph(h_\mu h)=\ch_\mu(h_\mu)\ph(h)\hst\fa h_\mu\in H_\mu.}
Thus we are in the same setting as that detailed in section \ref{Hqz}
and so we have a irreducible unitary representation $\pi_\mu$ of $H$
on $\calH_\mu$.

We can then define $E=N\x_H\calH_\mu$ (where
$(n,v)\sim(nh,\pi_\mu(h^{-1})v)$) and sections of this bundle can be
represented by functions $\ps:N\x H\to \Bbb C$ satisfying
(\ref{fncreln}). Further, by construction, these functions correspond
to polarized sections of $B$; hence we have proved
\begin{theorem}
There is a one-to-one correspondence
between the polarized sections of the prequantum bundle $B$ and the
sections of $E=N\x_H\calH_\mu$.
\end{theorem}

One advantage of identifying sections of the prequantum bundle $B$
with sections of $N\x_H\calH_\mu$ is that the latter bundle is closely
related to induced representations as we shall see in the next
section, but first we return to a matter alluded to at the end of
section \ref{preqbundle}.

{\emfl{The generalised Aharanov-Bohm effect}}\nopagebreak
Wu and Yang \ci{WY} gave a description of a generalised (i.e., the
gauge group is non-Abelian) Aharanov-Bohm effect in terms of a {\em
nonintegrable phase factor}. This is the ``generalised phase change''
of the wave function of the particle on being parallel transported
between two points with respect to the connection which represents the
gauge field. The term ``generalised phase change'' is used because the
nonintegrable phase factor acts via an irreducible representation of
the gauge group on the wave function of the particle. This
representation is, in general, not one-dimensional.

The Aharanov-Bohm effect for the gauge group SU(2) has been studied
\ci{HorvAB3} in terms of particles satisfying the Dirac equation, or
its non-relativistic limit, confirming Wu and Yang's predictions. In
the context of geometric quantization, the study of the generalised
Aharanov-Bohm effect has been constrained to trying to classify the
different ``prequantizations'' of the ``isospin'' bundle $N\to
N/H_\mu$ \ci{HorvABgq}. This was found only to be possible when the
relevant bundles were trivial. We will now show, by considering the
bundle $E=N\x_H\calH_\mu$ rather than the line bundle associated to
$N\to N/H_\mu$, how the Wu and Yang nonintegrable phase factor appears
naturally in our approach together with the representation of the
gauge group $H$ via which the phase factor acts on the wave function
of the particle.

The bundle $E=N\x_H\calH_\mu$ is an associated vector bundle of the
principal bundle $N\to N/H$. Now this latter bundle has a connection
\al and thus there is a corresponding covariant derivative,
$\del^\al$, on the sections of $E$. We will now show that this
covariant derivative is equivalent to the one on the sections of the
line bundle $B'=N\x_{H_\mu}\Bbb C$. This means it is permissible to
consider parallel transport in $E$ rather than in $B'$.

Let $s$ be a section of $E$. We can represent $s$ by
$s(q)=[n(q),\ps(n(q),h)]_H$ where \ps satisfies (\ref{fncreln}) and
$\pi_{N\to N/H}(n(q))=q\in Q=N/H$. Now consider a curve $\si(t)$ in
$Q$. We can choose $n$ so that $n(\si(t))=\tilde\si(t)$ is an
arbitrary horizontal lift of $\si(t)$ with respect to the connection
$\al$. Let $X$ be the tangent to $\si(t)$ at $t=0$. Then
\eq{\del^\al_Xs=\left[\tilde\si(0),\left.\frac{d}{dt}
\ps(n(\si(t)),h)\right|_{t=0}\right]_H\ll{del}.}

Now we saw earlier how sections of $E$ could be identified with
sections of $B'$. Here $s(q)$ corresponds to a section $s'$ of $B'$
where, with $\ga$ as defined in
(\ref{gadef}),
\eq{s'(q')=[n(q)h^{-1},\ga(n(q)h^{-1})]_{H\mu},}
and $q'=\pi_{N\to N/H_\mu}(n(q)h^{-1})$ . Let $\si'(t)=\pi_{N\to
N/H_\mu}(\tilde\si(t)h^{-1})$,
then
\eq{s'(\si'(t))=[\tilde\si(t)h^{-1},\ga(n(\si(t))h^{-1})]_{H_\mu}.}
It is easy to see that $\tilde\si(t)h^{-1}$ is a horizontal lift of
$\si'(t)$ with respect to $\al'$ since if $A$ is the tangent vector to
$\tilde\si(t)$ (so $\al(A)=0$), then
$\al'(\rh_{h^{-1}*}A)=pr(Ad_h(\al(A)))=0$. Thus, letting $X'$ be
the tangent vector to $\si'(t)$ at $t=0$, we have
\eq{\del^{\al'}_{X'}s'=\left[\tilde\si(0)h^{-1},\left.
\frac{d}{dt}\ga(n(\si(t))h^{-1})\right|_{t=0}\right]_{H\mu};}
and the right hand side corresponds to the section
$[\tilde\si(0),\frac{d}{dt}\ps(n(\si(t)),h)|_{t=0}]_H$ of $E$ in
agreement with (\ref{del}).

Now let $\si(t)$ be a curve in $Q$ with $\si(0)=q_0$ and
$\si(1)=q_1$. Denoting, as before, $\tilde\si(t)$ to be the horizontal
lift of $\si(t)$ to $N$ with respect to the connection $\al$, define
$h\in H$ by
\eq{\tilde\si(1)=\tilde\si(0)h.}
The general expression for $h$ is
\eq{h=P\exp\left(-\int_{q_0}^{q_1}\calA_adq^a\right)\ll{hdef2},}
where $P$ is a path-ordering operator along $\si(t)$ (which is
necessary as $H$ is, in general, no longer Abelian) and $\calA$ is the
local form on $Q$ for $\al$. This is the nonintegrable phase factor
of Wu and Yang \ci{WY}. For a section $s=[n,v]_H$ of $E$, the
change in $v\in\calH_\mu$ on $s$ being parallel transported around \si
is given by $\pi_\mu(h)$, i.e.,
\eq{v\to\pi_\mu\left(
P\exp\left(-\oint_\si\calA_adq^a\right)\right)v.\ll{genphase}}

When $\si(t)$ is closed, i.e., $q_0=q_1$, the ``phase change'' given
in (\ref{genphase}) is the generalised version of (\ref{phase}) and
is the corresponding Aharanov-Bohm effect for arbitrary $H$. (Note
that this result is in agreement with the case $H={\rm U}(1)$ considered in
section \ref{preqbundle} since when $H$ is Abelian $\pi_\mu=\ch_\mu$.)

\ssect{Induced Representations}
The theory of induced representations is well known for the case where
one starts with a representation $\pi_\mu$ of $H$ acting on
$\calH_\mu$ and induces, from $\pi_\mu$, a representation $\pi^\mu$
for a Lie group $G$ where $H\subset G$. The induced representation
$\pi^\mu$ acts on sections of the bundle $G\x_H\calH_\mu$. (E.g., see
\ci{Vara,BR}.) Now a generalisation of this type of induced
representation, due to Moscovici
\ci{Mosc}, exists for the case in hand of the bundle $N$. The starting
point is the bundle $E=N\x_H\calH_\mu$, given in the previous section,
with sections of $E$ identified with functions $\Ps:N\to\calH_\mu$
satisfying $\Ps(nh)=\pi_\mu(h^{-1})\Ps(n)$ for all $h\in H$. The
representation, $\pi^\mu$, of a group $G'$ which acts on the left on
$N$ {\em and} commutes with the right action of $H$ is given by
\eq{(\pi^\mu(g)\Ps)(n)=\Ps(g^{-1}n),\hst g\in G'.\ll{Autaction}}

We are naturally interested in taking $G'={\rm Aut\ }N$. (Note this is
not a special case of \ci{Mosc} since it was assumed there that $G'$
was locally compact.) We expect that, for $A\in\calL({\rm Aut\ }N)$,
the action given by $d\pi^\mu(A)$ corresponds to the observable $\hat
J_\mu(A,0)$ where $\hat J_\mu:\calL({\rm Aut\ }N\lx\cinf{Q})\to
\cinf{P_{\calO_\mu}}$ is defined just before (\ref{Jlocation}) and
$d\pi^\mu$ is defined, via (\ref{derivdef}), on the domain of
compactly supported cross-sections of the vector bundle $E$. Before
showing that this is the case, we remark that, as we will see, the
group \cinf Q can be incorporated into $\pi^\mu$ in an obvious way to
give a unitary representation of Aut $N\lx\cinf Q$. This
representation is the same as that used by Landsman \ci{LandSD} except
here the choice of such a representation is now fully justified in
that we show that the derived Lie algebra representation corresponds
to specific classical observables via the map $\hat J_\mu$. Also we
note that this representation of Aut~$N\lx\cinf Q$ essentially appears
in Isham
\ci[chapter 5.2]{Isham} under the guise of {\em lifted} group
actions. Isham starts with a group action on $Q$ and then considers
possible lifts of this to an automorphism of $N$. We have avoided the
use of such lifted actions by starting with the group Aut~$N$ to begin
with.

The action of Aut $N$ on a function \Ps given via (\ref{Autaction})
corresponds  to an action on the sections of the bundle
$B'=N\x_{H_\mu}\Bbb C$. Specifically these sections of $B'$ are
represented by functions $\ga:N\to\Bbb C$ satisfying
$\ga(nh_\mu)=\ch_\mu(h_\mu^{-1})\ga(n)$ where $h_\mu\in H_\mu$. The
action $\pi^\mu$ on these sections is then
$(\pi^\mu(\ph)\ga)(n)=\ga(\ph^{-1}n)$. In terms of a section, $s$, of
$B'$ we have
\eq{(\pi^\mu(\ph)s)(q')=\ph s(\bar\ph^{-1}_0q'),\hst q'\in N/H_\mu.\ll{Autrep}}
Here, $\bar\ph_0$ denotes the diffeomorphism defined on $N/H_\mu$
in the same fashion as in (\ref{diffeodef}). Returning to the
convention for projection maps of bundles used in section \ref{symleaves2},
sections $s$ then pullback to give sections $j^*s$ of the prequantum
bundle $B=N^\#\x_{H_\mu}\Bbb C$, where $j=\pi_{N/H_\mu}\crc \i$. Let
$\pi^\mu_0$ denote the corresponding action of $\pi^\mu$ on these sections,
i.e.,
$\pi^\mu_0(\ph)(j^*s)=j^*(\pi^\mu(\ph)s)$. Using
the realisation $N^\#=\{(n,p):n\in N,\ p\in T^*_{\pi_{N\to Q}(n)}Q\}$
and denoting an element of $N^\#/H_\mu$ by
$([n]_{H_\mu},p_{\pi_{N/H_\mu}(n)})$, we find
\eq{\ta^B_{\ph^{-1}}((\pi^\mu_0(\ph)(j^*s))([n]_{H_\mu},p_{\pi_{N\to
Q}(n)}))=(j^*s)([\ph^{-1}n]_{H_\mu},\bar\ph^*p_{\pi_{N\to
Q}(n)})\ll{flows}.}  Here $\ta^B$ denotes the left action of \ph on
elements of $B=N^\#\x_{H\mu}\Bbb C$ via
$\ta^B_\ph[\be_n,z]_{H_\mu}=[\ta_{(\ph,0)}\be_n,z]_{H_\mu}$, with
$\be_n\in N^\#\subset T^*N$. The vector field $V$, generated by the
infinitesimal action of $A\in\calL({\rm Aut\ }N)$ on $B$ via $\ta^B$,
intrinsically characterises the classical observable to which the
representation $\pi^\mu_0$ corresponds. We write $V=A^B$, where
the superscript $B$ denotes the space on which Aut $N$ is
acting. Recall, that in section \ref{prequantum}, we gave the relation
between an observable, its corresponding vector field on $B$ and the
resulting prequantum operator. (The term prequantum is used to
emphasise that these operators are regarded as acting on general
sections of $B$ rather than the polarized ones.) We intend to use this
relation to show
\begin{theorem}\ll{prop1}
The prequantum operator corresponding to the observable $\hat J_\mu(A,0)$ is
given by $\hbar d\pi_0^\mu(A)$.
\end{theorem}
To begin with we return to (\ref{flows}) and note that we can write this as
\eq{\ta^B_{\ph}((\pi^\mu_0(\ph^{-1})(j^*s))([n]_{H_\mu},p_{\pi_{N\to
Q}(n)}))=(j^*s)(\ta_{(\ph,0)}([n]_{H_\mu},p_{\pi_{N\to
Q}(n)})\ll{flows2}).} This now corresponds to (\ref{flowsdef}) since
$\xi_{\hat J_\mu(A,0)}=A^{P_\mu}$. It now remains to show that the
vector field $A^B$ corresponds to the observable $\hat J_\mu(A,0)$. The
verification of this result is technical and we first present two lemmas.
\begin{lemma}\ll{thm1}\mbox{}
\begin{enumerate}
\item[(i)]
\eq{\pi_{B\to P_\mu*}A^B=\xi_{\hat J_\mu(A,0)},\ll{observcond1}}
\item[(ii)]
\eq{\hbar\langle \tilde\Th,A^B\rangle =\hbar\langle
\bar{\tilde\Th},A^B\rangle
=-(\hat J_\mu(A,0))\crc\pi_{B\to P_\mu}.\ll{observcond2}}
\end{enumerate}
\end{lemma}
{\it Proof.}
Now $\pi_{B\to P_\mu*}A^B$ is just the vector field generated by
$A$ acting on $P_\mu$. Hence, from the properties of the momentum map
$\bar J_\mu$, it is evident that (\ref{observcond1}) holds.

To verify (\ref{observcond2}), let $b:U\subset
N/H_\mu\to N$ denote a section of the bundle $N\to N/H_\mu$. Then the
section $j^*b$ gives a local trivialization \ta of $B$ via
$\ta(q',z)=[(s^*b)(q'),z]_{H_\mu}$. In this trivialization the
connection one-form is given by
\eq{\tilde\Th=\hbm\pi^*_{B\to P_\mu}\Th-i\ta^{-1*}\frac{dz}{z},\ll{conntriv}}
where \Th is the local potential one-form of the connection given in
(\ref{oneform}) (with the minus sign as we are using a right action of
$H$). From the definition of $\th_0$ in (\ref{canoneform}), and
recalling that elements $(n,p)$ of $N^\#$ correspond to elements in
the annihilator of the vertical bundle of $TN$, we have
\eq{\langle \pi^*_{B\to P_\mu}\i^*\th_0,A^B\rangle
_{[(n,p),z]_{H_\mu}}=\langle \pi^*_{N\to
Q}p,A^N\rangle _n.\ll{term0}}

The remaining part of $\Th$, as given in (\ref{oneform}), consists of
an $\al_\mu$ term. Now
\eq{\langle \pi^*_{B\to P_\mu}j^*b^*\al_\mu,A^B\rangle _{[(n,p),z]_{H_\mu}}=
\langle \pi^*_{N\to N/H_\mu}b^*\al_\mu,A^{B'}\rangle _n;\ll{term1}}
further,
\eq{\langle \ta^{-1*}\frac{dz}{z},A^B\rangle _{[(n,p),z]_{H_\mu}}=
\langle \ta_1^{-1*}\frac{dz}{z},A^{B'}\rangle _{[n,z]_{H_\mu}},\ll{term2}}
where $\ta_1$ is the local trivialization of $N\x_{H_\mu}\Bbb C$ via
$\ta_1(q,z)=[b(q),z]_{H_\mu}$. To complete the verification of
(\ref{observcond2}) we use the following result.
\begin{lemma}
\eq{\langle \tilde\al_\mu,
A^{B'}\rangle _{[n,z]_{H_\mu}}=\langle \bar{\tilde\al}_\mu,
A^{B'}\rangle _{[n,z]_{H_\mu}}=
\hbm\langle \al_\mu,A^N\rangle _n,\ll{connid}}
where $\tilde\al_\mu=\hbm\pi^*_{N\to
N/H_\mu}b^*\al_\mu-i\ta_1^{-1*}\frac{dz}{z}$ is the connection
one-form on $B'$ and $\bar{\tilde\al}_\mu$ is its complex conjugate.
\end{lemma}
{\it Proof.} This can be readily checked by considering, for
example, the curve $b(q(t))e^{tA'}$ in $N$ and the
corresponding curve $[b(q(t))e^{tA'},z]_{H_\mu}$ in $B'$; here
$A'\in\h_\mu$. $\blacksquare$

Thus, it finally remains to calculate $\langle \al_\mu,A^N\rangle _n$. We
have
\eq{\langle \al_\mu,A^N\rangle _{s(q)h}=\langle
\cod{h}\mu,\al_{s(q)}(A^N)\rangle ,}
where $s$ is the section of the bundle $N\to N/H$ used at the end of
section \ref{observqz}.
Combining this equation with (\ref{term0}) we finally obtain
\eq{\hbar\langle \tilde\Th,V\rangle _{[(s(q)h,p),z]_{H_\mu}}=-\langle
\pi^*_{N\to
Q}p,A^N\rangle _{s(q)} - \langle \cod{h}\mu,\al_{s(q)}(A^N)\rangle .}
Note that the right hand side is in fact a function on $P_\mu\iso
P_{\calO_\mu}=N^\#\x_H\calO_\mu$.
Now for $[\be_{s(q)},\nu]_H\in N^\#_{s(q)}\x_H\calO_\mu$ we can write
this as $[\rh^*_{h^{-1}}\be_{s(q)},\mu]_H\in N^\#_{s(q)h}\x_H\{\mu\}$
where $h$ is such that $\cod{h}\mu=\nu$. So we can write
\eqn{\hbar\langle \tilde\Th,V\rangle _{[[\be_{s(q)},\nu]_H,z]_{H_\mu}}&=&
-\langle \pi^*_{N\to Q}p,A^N\rangle _{s(q)}-\langle
\nu,\al_{s(q)}(A^N)\rangle \nn\\ &=&-(\hat
J_\mu(A,0))[\be_{s(q)},\nu]_H}
using (\ref{globalobserv}). Clearly the
same expression is obtained for $\hbar
\langle\bar{\tilde\Th},V\rangle$ and thus this completes the proof of
Lemma \ref{thm1}. $\blacksquare$

By considering (\ref{flows2}), (\ref{qopdef}) and Lemma \ref{thm1}
we see that Theorem \ref{prop1} has been proved.

It finally remains to incorporate $\cinf{Q}$ into the representation
$\pi^\mu$ defined in (\ref{Autrep}). For $(\ph,f)\in {\rm Aut\
}N\lx\cinf{Q}$, following \ci{LandSD} we put
\eq{(\pi^\mu(\ph,f)s)(q')=e^{-i\hbm f(\pi_{N/H_\mu \to Q}(q'))}\ph
s(\bar\ph_0^{-1}q'),}
which gives a [left] representation of Aut $N\lx\cinf{Q}$ on the
sections of $B'$. Note that, in terms of the functions \Ps used in
(\ref{Autaction}), the representation $\pi^\mu$ is given by
\eq{(\pi^\mu(\ph,f)\Ps)(n)=e^{-i\hbm f\crc\pi_{N\to
Q}(n)}\Ps(\ph^{-1}n).\ll{therep}}
Concentrating on the case $\pi^\mu(0,f)$ we claim
\begin{theorem}\ll{prop2}
The prequantum operator corresponding to the observable $\hat
J_\mu(0,f)$ is given by $\hbar d\pi_0^\mu(0,f).$
\end{theorem}
Identifying the Lie algebra of $\cinf{Q}$ with the Lie group we have,
in the notation of (\ref{flowsdef}),
$\rh_t[\be_n]_{H_\mu}=[\be_n-t\pi^*_{N\to Q}df]_{H_\mu}$. Now let
$\hat\rh_t=\pi_0^\mu(0,-tdf)$, i.e., in a local trivialization of $B$
\eq{(\hat\rh_t j^*s)([n]_{H_\mu},p)=([n]_{H_\mu},p,e^{i\hbm
tf\crc\pi_{N\to Q}(n)}\ps([n]_{H_\mu})),} where $s$ is a section
determined locally by $\ps\in\cinf{U\subset N/H_\mu}$, i.e.,
$(j^*s)([n]_{H_\mu},p)=([n]_{H_\mu},p,\ps([n]_{H_\mu}))$. It
follows that the corresponding $\de_t$ is given by
\eq{\de_t((j^*s)([n]_{H_\mu},p))=([n]_{H_\mu},p-tdf,e^{-i\hbm
tf\crc\pi_{N\to Q}(n)}\ps([n]_{H_\mu})).}
Regarding, as before, $\de_t$ to be the flow of the vector field
$V=\xi^B(f)$ (i.e., the vector field on $B$ generated by the action of
$f$ via $\ta$) we  now only need to prove
\noindent
\begin{lemma}\mbox{}
\begin{enumerate}
\item[(i)]
\eq{\pi_{B\to P_\mu*}\xi^B(f)=\xi_{\hat J_\mu(0,f).};\ll{observcond3}}
\item[(ii)]
\eq{\hbar\langle \tilde\Th,\xi^B(f)\rangle=
\hbar\langle \bar{\tilde\Th},\xi^B(f)\rangle=
-f\crc\pi_{B\to
Q}.\ll{observcond4}}
\end{enumerate}
\end{lemma}
{\it Proof.} As in Lemma \ref{thm1}, (\ref{observcond3}) follows
from the properties of the momentum map $\bar J_\mu$.  Also it is easy
to see from (\ref{canoneform}) that $\langle \pi^*_{B\to
P_\mu}\i^*\th_0, \xi^B(f)\rangle =0$. Thus it just remains to calculate
$\langle \pi^*_{B\to P_\mu}j^*b^*\al_\mu,\xi^B(f)\rangle $. Using
(\ref{conntriv}) we obtain
\eq{\hbar\langle \tilde\Th,\xi^B(f)\rangle _{[(n,p),z]_{H_\mu}}=
\hbar\langle \bar{\tilde\Th},\xi^B(f)\rangle _{[(n,p),z]_{H_\mu}}=
-f\crc\pi_{N\to
Q}(n),}
 as required by (\ref{observcond4}). $\blacksquare$

In the same manner as Theorem \ref{prop1} this completes the proof
of Theorem \ref{prop2}. Note that $(\hbar
d\pi^\mu_0(0,f)\ps)(q')=f(\pi_{P_\mu\to Q}(q'))\ps(q')$ as expected.

So far we have found operators which act on the sections of $B$. We
must now restrict these to act on {\em polarized} sections of
$B$. Recall that there is a one-to-one correspondence between the
polarized sections of $B$ and the sections of
\mbox{$E=N\x_H\calH_\mu$}. Thus, we see that the action of $\pi^\mu_0$ on the
polarized sections is equivalent to the action of $\pi^\mu$ on the
sections of $E$. So, to summarize, we can put our results for Aut~$N$
and $\cinf{Q}$ together to obtain our key result: the prequantum operator
corresponding to the classical observable $\hat J_\mu(A,f)$ is given
by $\hbar d\pi^\mu(A,f)$. However, this is not quite the complete
picture because so far we have considered the quantum operators to be
acting on sections of $B$ rather than sections of $B_K=B\otimes\de_D$.
We address this point in the next section.

\ssect{Unitary representations}
As it stands the representation $\pi^\mu$ fails to be unitary unless
there exists a measure on $N$ which is invariant under Aut $N$. This
can be overcome in a standard way using the Radon-Nikodym derivative,
e.g., \ci[chapter 5.2]{Isham}. Let \muu be an $H$-invariant
measure on $N$, which in turn determines a measure $\nu$ on $Q$. Then
we define the representation $\pi^\mu$ by replacing (\ref{therep})
with
\eq{(\pi^\mu(\ph,f)\Ps)(n)=\left(\frac{d\mu(\ph^{-1}n)}{d\mu(n)}\right)^{1/2}
e^{-i\hbm f\crc\pi_{N\to Q}(n)}\Ps(\ph^{-1}n).\ll{therepnew}}
This
then gives a unitary representation of Aut $N\lx C^\infty_c(Q)$, where
$C^\infty_c(Q)$ is the subspace of smooth functions on $Q$ with
compact support. The addition of the square-root term corresponds to
the replacement of $B$ by $B_K$ in section \ref{gengq} and the fixing
of a choice of \et such that $\et\bar\et=d\nu$. The inner product is
(c.f.\ \ref{innerp})
\eq{(\Ps,\Ps')=\int_Q d\nu(\pi_{N\to Q}(n))\
(\Ps(n),\Ps'(n))_{\calH_\mu},}
and we restrict our attention to smooth functions \Ps that have
compact support.

Further the representation $\pi^\mu$ is, in general, irreducible. To
see this we first recall that there is a one-to-one correspondence
between the sections of $E$ and the functions $\ga:N\to\Bbb C$ which
satisfy (\ref{preqcond}) and the ``pseudo-polarization'' condition
detailed in section \ref{poldsections}. In terms of the $\ga$'s we
have
\eq{(\pi^\mu(\ph,f)\ga)(n)=\left(\frac{d\mu(\ph^{-1}n)}{d\mu(n)}\right)^{1/2}
e^{-i\hbm f\crc\pi_{N\to Q}(n)}\ga(\ph^{-1}n).\ll{pirep3}}
For a general Lie group $G$, the representation $\hat\pi$ of
$G\lx\cinf N$ acting on smooth functions $\ps:N\to\Bbb C$, given by
\eq{(\hat\pi(g,f)\ps)(n)=\left(
\frac{d\mu(g^{-1}n)}{d\mu(n)}\right)^{1/2}e^{-if(n)}\ps(g^{-1}n)}

is irreducible \ci{Isham} provided $N$ does not decompose into a
disjoint union of two $G$-invariant subsets both of which have
positive $\mu$-measure. (This situation can be overcome if \muu is
required to be $G$-ergodic.) Returning to our representation $\pi^\mu$
we see that it is closely related to $\hat\pi$ except that firstly we are
considering a subspace of $\cinf{N,\Bbb C}$, i.e., the functions \ga
satisfying the conditions noted above. This does not alter the
irreducibility of $\hat\pi$. Secondly, the function $f$ in
(\ref{pirep3}) is lifted to one on $N$ which means that it cannot vary
along the fibres of $N\to N/H$. However, this restriction merely
ensures that the action of $\pi^\mu$ is to create a function which
satisfies the conditions on the $\ga$'s. Hence $\pi^\mu$ is
irreducible, provided $N$ does not decompose into a disjoint union of
two Aut~$N$ invariant subsets.

Landsman \ci{LandSD} has given an explicit form of the representation
$\pi^\mu$ in terms of functions $\ps^\al:U_\al\subset Q\to\calH_\mu$.
Specifically, cover $Q$ with open sets $\{U_\al\}$ and denote local
smooth sections of $N$ by $s_\al:U_\al\to N$ such that for $q\in
U_\al\cap U_\be$ $s_\be(q)=s_\al(q)g_{\al\be}(q)$ where
$g_{\al\be}:U_\al\cap U_\be\to H$ is the transition function for the
two coordinate patches $U_\al$ and $U_\be$. An element of $\calH^\mu$
is represented by a collection $\{\ps_\al\}$ of smooth functions
$\ps_\al:U_\al \to \calH_\mu$, which are related on $U_\al\cap U_\be$
by
\eq{\ps_\al(q)=\pi_\mu(g_{\al\be(q)})\ps_\be(q)\ll{origpsilink}.}
The action of $\pi^\mu$ on these functions is given by
\eq{(\pi^\mu(\ph,f)\ps_\al)(q)=\left(\frac{d\nu(\bar\ph^{-1}q)}
{d\nu(q)}\right)^{1/2} e^{-i\hbm
f(q)}\pi_\mu((h_\be[\ph^{-1}(s_\al(q))])^{-1})\ps_\be(\bar\ph^{-1}q),}
where $h_\be$ is the element of $H$ satisfying
$s_\be(\bar\ph^{-1}q)h_\be= \ph^{-1}(s_\al(q))$. Here it is assumed
that $q\in U_\al$ and $\bar\ph^{-1}q\in U_\be$. Landsman \ci{LandSD}
also gives a
formula for the derived representation $d\pi^\mu$, which in our
notation is
\eqn{\hbar(d\pi^\mu(A,f)\ps_\al)(q)&=&f(q)\biggl(-i\hbar\left[\del_{\pi_{N\to
Q*}A}+\frac{1}{2} {\rm div\ } (\pi_{N\to
Q*}A)(q)\right]\nn\\
&&\mbox{}+d\pi_\mu(\al_{s_\al(q)}(A))\biggl)\ps_\al(q). \ll{derivrep}}
Here $\al$ is the connection on $N\to Q$ and \del is the corresponding
covariant derivative via the representation $\pi_\mu$ of $H$ on
$\calH_\mu$. Also note that the div term agrees with that in
(\ref{ftildeaction}). As noted by  Landsman\ci{LandSD} the operator
$d\pi^\mu$ is defined and essentially self-adjoint on the domain of
compactly supported cross-sections of the bundle $E=N\x_H\calH_\mu$.
Further, the right hand side of (\ref{derivrep}) is actually
independent of the connection used. The motivation for writing
(\ref{derivrep}) in this manner is that the third term on the right
hand side is the generalisation of the Poincar\'e term in the angular
momentum of a charged particle moving in the field of a magnetic
monopole \ci{Landhom}; if $A$ is a symmetry of the dynamics then
this term is the contribution of the external gauge field to the
conserved operator $d\pi^\mu(A,0)$.

We have now proved our main result:
\begin{theorem}
For the constrained mechanical system whose reduced phase space is
$P_\mu$, the quantum operator corresponding to the classical
observable $\hat J_\mu(A,f)$ is given by $\hbar d\pi^\mu(A,f)$ and
acts on  compactly supported cross-sections of the bundle
$E=N\x_H\calH_\mu$. In terms of the quantizing map $Q_\hbar$, this is
written as
\eq{Q_\hbar(\hat J_\mu(A,f))=\hbar d\pi^\mu(A,f)\ll{theqzmap}.}
\end{theorem}
Note that as $\pi^\mu$ is, in general, irreducible we have satisfied
the irreducibility requirements discussed in section \ref{intro}.

We now turn to the problem of finding the Hamiltonian for the quantum
system. Unfortunately the classical Hamiltonian is not in the subclass
of observables that we can quantize using $Q_\hbar$; indeed this is a
generic problem with geometric quantization when the Hamiltonian is
not linear in momentum. However, Landsman \ci{LandSD} has shown that
the quantum Hamiltonian, $H_\hbar$, is given by the gauge-covariant
Laplacian on $E=N\x_H\calH_\mu$. Specifically, $H_\hbar$ determines
the time-evolution of an operator $\hat L=Q_\hbar(\hat J_\mu(A,f))$ via
\eq{\hat L(t)=e^{it\hbm H_\hbar}\hat Le^{-it\hbm H_\hbar},}
and $H_\hbar$, which acts on sections of $E$, is given by
\eq{H_\hbar=-\frac{1}{2}\hbar^2\del\cdot\del+V_0.}
Here we have included the potential $V_0$, which was defined at the end
of section \ref{observqz}, and we note that the gauge-covariant
Laplacian is defined with respect to the connection $\al$.

Locally, we can use
$(h^1,\ldots,h^{d_H},q^{d_H+1},\ldots,q^{d_N})$ as coordinates on $N$, where
$(q^{d_H+1},\ldots,q^{d_N})$ are  coordinates on $Q$ and
$(h^1,\ldots,h^{d_H})$ are coordinates on the fibre $H$. We can motivate
$H_\hbar$ as the Hamiltonian if the coordinates are chosen such that
$\al(\cdd{}{q^\al})=0$ and $d\nu=d^nq$. The latter condition means
that det~${\sf g}=1$, where {\sf g} is the metric on $Q$. In the
notation of (\ref{Ham3}), we then find that $H_\hbar$ coincides with
$H'_\hbar$, where
\eq{H'_\hbar=-\frac{1}{2}\hbar^2Q_\hbar(\hat J_\mu(A_\al,0)){\sf
g}^{\al\be} Q_\hbar(\hat J_\mu(A_\be,0))+\hat J_\mu(0,V_0).}
Note that $\Bbb I^{IJ}Q_\hbar(\hat J_\mu(A_I,0))Q_\hbar(\hat
J_\mu(A_J,0))=\Bbb I^{IJ}d\pi_\mu(T_I)d\pi_\mu(T_J)$ is a Casimir
operator for $H$ and, as the representation $\pi_\mu$ is irreducible,
this is a constant which can be omitted from the Hamiltonian. Hence
$H'_\hbar$ can be considered to be the quantum operator corresponding
to $H_{\calO_\mu}$ given in (\ref{Ham3}). To see that $H_\hbar$ agrees
with $H'_\hbar$ it is easier to use the representation $\pi^\mu$ as
defined on functions \Ps used in (\ref{Autaction}). Then the action of
$d\pi^\mu$ is given by \ci{LandSD}
\eq{\hbar (d\pi^\mu(A,0)\Ps)(n)=-i\hbar\left(\left(A+\frac{1}{2}{\rm div\
}A\right)\Ps\right)(n).}
Thus, noting that div~$\cdd{}{q^\al}=0$, we find
\eq{H'_\hbar=-\frac{1}{2}\hbar^2\cdd{}{q^\al}{\sf
g}^{\al\be}(q)\cdd{}{q^\be},}
whilst, up to a constant, $H_\hbar$ acting on the functions \Ps is given
by $-\frac{1}{2}\hbar^2\De_{\rm LP}$, where $\De_{\rm LP}$ is the
Laplace-Beltrami operator. Thus, in this choice of coordinate system,
$H_\hbar$ agrees with $H'_\hbar$.

\ssect{Homogeneous spaces}
\ll{homspaces}
When the bundle $N$ is a finite-dimensional Lie group $G$ (with
$H\subset G$) the configuration space $Q=G/H$ is homogeneous. Isham
\ci{Isham} has considered quantization on such configuration spaces
and in this section we relate our work to his. In particular we can
explain two unresolved features of Isham's method.  The first is the
appearance of {\em inequivalent quantizations}, i.e., the discovery of
many different quantum schemes resulting from the same classical
system. The second is the presence of quantizations which appear to be
unrelated to the original system. We find that the geometric
quantization approach shows that the inequivalent quantizations of
Isham's correspond to slightly different classical systems, and also
that the seemingly unrelated quantizations of Isham's are indeed
quantizations resulting from a completely different physical system.
Note that the different quantizations Isham finds are unrelated to
whether or not the configuration space is multiply connected.  We
conclude the section with a worked example for the case $G={\rm
SU}(2)$ and $H={\rm U}(1)$. This gives the homogeneous configuration
space $S^2$.

Let $V$ be a vector space which carries an almost faithful
representation of $G$ and for which there is a $G$-orbit in $V$ that
is diffeomorphic to $G/H$. Isham \ci{Isham} argues that quantization
corresponds to representations of the semi-direct product group
$\calG=G\lx V^*$. Crucially Isham considers $\calG$ as a subgroup of
\mbox{${\rm Diff\ }Q\lx\cinf Q/\Bbb R$} (where $\Bbb R$ denotes the
functions constant on $Q$) and the phase space of the system to be
$T^*Q$. A momentum map for the action of $\calG$ on $T^*Q$ is found
and indeed corresponds to the restriction of the momentum map $\bar
J_{\mu=0}$ of section \ref{observqz} to $\calG\subset {\rm Aut\ }N\lx\cinf
Q$. Isham quantizes the system by finding irreducible unitary
representations of \calG (via Mackey theory) and using the momentum
map to match observables on $T^*Q$ with the generators of the
representations of $\calG$.

We can split the irreducible unitary representations of \calG into two
classes, those which arise from consideration of a $G$-orbit $\Th\subset V$
where $\Th\iso G/H$ (the first class) and those from a $G$-orbit
$\Th'\subset V$ where $\Th'\iso G/H'$ with $H\not\simeq H'$ (the
second class). We can now compare Isham's results to our own.
Specifically the representations $\pi^\mu$ we find are the same as
those in Isham's first class. (Here we are restricting $\pi^\mu$ to
$\calG$.) Crucially, however, each of our representations corresponds,
via $\mu$, to a different symplectic leaf in $(T^*G)/H$. Further,
each symplectic leaf has a different momentum map and thus each of the
different representations corresponds to a slightly different physical
system. In terms of the quantizing map $Q_\hbar$, Isham considers the
phase space $G^\#/H\iso T^*Q$ with
\eq{Q_\hbar(\hat J_{\mu=0}(A,u))=\hbar d\pi^\mu(A,u),}
where $(A,u)\in\calL(\calG)^*\iso\g\x V^*$.
Note that it is not clear which representation $\pi^\mu$ is to be
chosen on the right hand side. Whereas we have the phase space
$G^\#\x_H\calO_\mu$ with
\eq{Q_\hbar(\hat J_{\mu}(A,u))=\hbar d\pi^\mu(A,u).}
It is now clear that different representations of \calG correspond to
different physical systems. In fact, for a particle moving in a
Yang-Mills field, the different representations of \calG correspond to
the different possible charges that the particle could have.

The representations of $\calG$ in Isham's second class clearly
correspond to the quantizations of constrained systems which have $H'$
as the symmetry (gauge) group. In terms of a particle  in a
Yang-Mills field, these representations correspond to a particle on
the configuration space $G/H'$ where the internal charge couples
to the gauge group $H'$. Thus, they are unrelated to the original system.

\sssect{The canonical connection}

There is a natural choice for the metric on $G$; specifically, as $H$
is compact, \h is reductive in \g and a positive definite inner
product $\langle \langle \ ,\ \rangle \rangle $ exists on \g which is
invariant by $\pi_{ad}(H)$ (e.g., \ci{KNII}). Thus, by defining \m to
be the orthogonal complement to $\h$, with respect to this inner
product, we have that $\g=\h\oplus\m$ and $[\h,\m]\subset\m$, i.e., the
decomposition is reductive. We can use this inner product on \g to
define one on $T_gG$ via
\mbox{$\langle \langle X,Y\rangle \rangle _g=$}\mbox{$\langle \langle
\la_{g^{-1}*}X,\la_{g^{-1}*}Y\rangle \rangle $}, thus defining a
metric on $G$. We saw in section \ref{Hsystem} that a choice of a
metric on $N$ was equivalent to choosing a connection on $N\to
N/H$. In this section we will explicitly identify this connection.

Let ${\sf g}_{ab}=\langle \langle T_a,T_b\rangle
\rangle $ where $\{T_i\}$ is a basis for \g. We can write the Hamiltonian as
\eq{H_0(g,p)=\frac{1}{2}{\sf g}^{ab}p_ap_b+V(g),}
where ${\sf g}^{ab}{\sf g}_{bc}=\de^a_c$ and the $\{p_i\}$
are coordinates on $T^*_gG$ in the left trivialization
(\ref{lefttriv}). The corresponding Legendre transformation
$\Bbb FL$ is $(g,v^j)\to(g,p_j)$ where $p_j={\sf g}_{ja}v^a$ and
$(g,v)\in G\x\g$ represents $\la_{g*}v\in T_gG$.

We denote the momentum map for the right action of $H$ on $T^*G$ by
$J_H$. From (\ref{genmommap}) we find for, $X\in\h$,
\eq{\langle J_H(g,p),X\rangle=\langle p,X\rangle.}
So $J_H^I(g,p)=p^I$ where $I=1,\ldots,d_H$. From the definition of
$\Bbb I$ (\ref{LIN}) we have $\Bbb I(g)_{IJ}={\sf g}_{IJ}$. To
calculate $\al^I(g,v)$, note that the choice of a reductive
decomposition means that $g^{I\be}=0$, for
$\be=d_H+1,\ldots,d_G$. Thus, we readily find
\eq{\al^I(g,v)=v^I\ll{alresult}.}
We see that
\al is the {\em canonical connection} on $G\to G/H$. The canonical
connection, $\om$, for this bundle is defined to be the \h component
of the canonical (Maurer-Carter) one-form on $G$ with respect to the
decomposition $\g=\h\oplus\m$ (e.g., see
\ci{KNI}).  Explicitly, $\om=T_I\otimes\th^I$, where
$\{\th^I\}$ are left invariant one-forms as defined in section
\ref{dualpairs} using a basis of \gs which is dual to the basis
$\{T_i\}$ of $\g$, i.e., $\langle d^a,T_b\rangle=\de^a_b$. If
$(g,v)_L\in G\x\g$ represents the point $\la_{g*}v\in T_gG$ then we
have $\om(\la_{g*}v)=T_Iv^I$. I.e., $\om^I=v^I$, in agreement with
(\ref{alresult}). The one-form $\al_\mu$ on $G$ is then just
\eq{\al_\mu=\mu_I\th^I.}

It is easy to see that the trivialization of $T^*G\iso G^\#\x\hs$
induced by (\ref{TSNtriv}) corresponds to the left trivialization of
$T^*G$ in (\ref{lefttriv}) where $G^\#\iso G\x\ms$. This allows us to
rewrite (\ref{THEmommap}) as
\eq{\langle J(g,p),(X,u)\rangle=\langle \cod gp,X\rangle+\langle
u,\si(g)a\rangle ,} where $(X,u)\in\calL(\calG)\iso\g\x V^*$ and \si
is the representation of $G$ on $V$. We regard $J$ as a momentum map
for the left action of $\calG$ on $T^*G$. In passing we note that as
\calG is finite-dimensional, $\calL(\calG)^*$ is foliated by
symplectic leaves which are the coadjoint orbits. If the explicit form
for the coadjoint action of $\calG$ on $\calL(\calG)^*$ is considered,
(e.g., see \ci{MRW}) it is easy to see that the coadjoint orbit
$\cod\calG (\nu,a)$, where $(\nu,a)\in \gs\x V$, is contained in
$(\gs,\si(G)a)$ which is exactly $J(T^*G)$. Hence the symplectic
leaves in $J(T^*G)$ are coadjoint orbits. Thus, by the arguments of
section \ref{observqz}, $\bar J_\mu$ is a symplectic diffeomorphism
which maps the symplectic leaf $G\x_H(\ms\x\calO_\mu)$ to the
symplectic leaf $\cod\calG (\nu,a)$ where $\nu\in\gs$ such that
$\nu\rst\h=\mu\in\hs$. Explicitly,
\eq{\bar J_\mu[g,p]_H= (\cod gp,\si(g)a)\in\calL(\calG)^*\iso\gs\x
V,\ll{themommap}}
where $[g,p]_H\in G\x_H(\ms\x\calO_\mu)$. This gives an elegant
alternative proof of a previously known result \ci{MRW}.

\sssect{The case $G={\rm SU}(2)$, $H={\rm U}(1)$}
\ll{iqS2example}
We can parametrize ${\rm SU(2)}$ using the Euler angles
$(\ph,\th,\ch)$:
\eq{g(\ph,\th,\ch)=e^{-\ph a_3}e^{-\th a_2}e^{-\ch a_3},}
where $a_j=1/2i\si_j$, the $\{{\si_j}\}$ are the Pauli spin matrices,
$0\leq\th\leq\pi$, $0\leq\ph\leq 2\pi$ and $0\leq\ch\leq 4\pi$. We
regard $H={\rm U(1)}$ as the subgroup
\eq{H=\{g(0,0,\ch):0\leq\ch\leq 4\pi\}.}
There is a standard homomorphism (e.g., see \ci{CornI})
$\al:{\rm SU(2)}\to {\rm SO(3)}$ given by
$\al(u)_{jk}=1/2{\rm tr}(\si_ju\si_ku^{-1})$.
In the parametrization above
this gives
\eq{G(\ph,\th,\ch) =\al(g(\ph,\th,\ch))=e^{-\ph A_3}e^{-\th
A_2}e^{-\ch A_3}.\ll{SO3rep}} Here $[A_i]_{jk}=-\varep_{ijk}$ and
(\ref{SO3rep}) is the standard Euler angle parametrization of ${\rm
SO(3)}$. The subgroup ${\rm U(1)}\subset{\rm SU(2)}$ is mapped by \al
to the subgroup $G(0,0,\ch)\iso {\rm SO(2)}$ whilst the kernel of \al
is just $\pm 1\subset {\rm U(1)}$. Hence we can see that \al drops to
a map on the quotient spaces ${\rm SU(2)}/{\rm U(1)}\to {\rm
SO(3)}/{\rm SO(2)}$.  This is readily observed to be a diffeomorphism;
thus ${\rm SU(2)}/{\rm U(1)}\iso S^2$. We can use \al to give a
representation of SU(2) on $\Bbb R^3$. Clearly the orbits of SU(2) in
$\Bbb R^3$ are then spheres (or just the origin).  Thus by choosing
$G$=SU(2), $H$=U(1) and $V=\Bbb R^3$ so that $\calG=$SU(2)$\lx \Bbb
R^{3*}$ we satisfy Isham's requirements for $\calG$. We use the
measure $d\nu=\sin\th d\ph\w d\th$ on $S^2$. Note that this is
$G$-invariant.

It is a standard result that the derived
map $\al':\su\to\so$ is an isomorphism with $\al'(a_i)=A_i$, e.g., see
 \ci{CornII}. Further, the adjoint action, $\pi_{ad}$, of SO(3) on
$\so\iso \Bbb R^3$ is the usual one, i.e., using
 $\{A_i:i=1,2,3\}$ as a basis for \so then if $p\in\Bbb R^3\iso\so$
and $g\in{\rm SO(3)}$ then $\pi_{ad}(g)\cdot p=g\cdot p$, e.g., see
\ci{AM}.
Similarly regarding $\so^*\iso\R^3$, we have
$\cod{g}p=g\cdot p$.
We can now write down an explicit expression for (\ref{themommap}).
 We have
\eq{\bar J_\mu[g,p]_H=(\al(g)\cdot p,\al(g)\cdot a_0),\ll{classlink}}
where we have used the fact that $\al'(\cod{g}A)=\cod{\al(g)}\al'(A)$.
Here $a_0=(0,0,1)^T$ so, as is standard, we are taking $S^2$ to have
unit radius.

We can use (\ref{derivrep}) to find explicit expressions for the
derived representation $d\pi^\mu$ where we are considering $\pi^\mu$
as representation for $\calG\subset{\rm Aut\ }N\lx\cinf Q$. Of course,
however, we must first find the representation $\pi_\mu$ of $H$ which
corresponds to the coadjoint orbit $\calO_\mu\subset \hs$.
Since $H$ is
Abelian the coadjoint orbits are just single
points; $\calO_\mu=\{\mu\}$. Finding the representation of $H$ which
corresponds to the orbit $\calO_\mu$
is a trivial example of the standard problem discussed in
section \ref{Hqz}, since $H_\mu$, the
isotropy group of $H$, is just $H$. Writing $X_\mu$
for the homomorphism $\ch_\mu$ of section \ref{Hqz}, then, if the
elements of $H_\mu$ are written as $e^{-\ch a_3}$, $0\leq
\ch\leq 4\pi$, we have from (\ref{chidef2})
\eq{X_\mu(e^{-\ch a_3})=e^{i\hbm\mu\ch},\hst 2n\in\Bbb Z,}
where we are regarding $\hs\iso\Bbb R$. Requiring $X_\mu$ to be a
single-valued function on $H_\mu$, we see that
the integrality condition is
\eq{\mu=n\hbar,\hst 2n\in \Bbb Z.}
The sections of the prequantum line bundle $B\to \calO_\mu$ are
identified with functions $\ps:H\to\Bbb C$ such that
\eq{\ps(hh')=X(h)\ps(h'),\hst h,h'\in H,}
and we obtain a representation $\pi_\mu$ of $H$ on the sections of $B$
by pulling
back the $\ps$'s under right translation. In this trivial case we can,
as $H_\mu=H$, identify each \ps with some $z\in \Bbb C$ via $\ps(e)=z$
so
$(\pi_\mu(h)\ps)(h')=\ps(h'h)=X(h)\ps(h')$.
Hence
\eq{\pi_\mu(e^{-\ch a_3})=e^{in\ch}.}
Thus, as expected, $\pi_\mu$ is an irreducible unitary representation
of $H$. In this simple case the process of finding a polarization does
not arise since $\calO_\mu$ is just a single point.

We now turn to finding the induced representation $\pi^\mu$.  We can
 cover $S^2$ with the standard coordinate patches
\eqn{M_N&=&\{(\ph,\th)\in S^2:0\leq\th< \pi/2+\ep,0\leq\ph\leq 2\pi\};\\
M_S&=&\{(\ph,\th)\in S^2:\pi/2-\ep <\th\leq 2\pi,0\leq\ph\leq 2\pi\},}
where $\pi/2 > \ep>0$ and $(\ph,\th)$ are the standard spherical polar
angles.  Following Landsman \ci{LandII}, we choose continuous sections
$s_{N/S}:M_{N/S}\to {\rm SU(2)}$, namely
\eqn{s_N(\ph,\th)&=&g(\ph,\th,-\ph)\ll{sdefN};\\
s_S(\ph,\th)&=&g(\ph,\th,\ph)\ll{sdefS},} so that $s_N$ is continuous
at the North pole ($\th=0$) while $s_S$ is at the South pole
($\th=\pi$). On the overlap region $M_N\cap M_S$
\eq{s_S(\ph,\th)=s_N(\ph,\th)e^{-2\ph a_3}\ll{slink}.}
Before calculating the explicit form for the induced representations
we first find an expression for the classical observables in local
coordinates on each coordinate patch. Note that the symplectic leaf
$P_{\calO_\mu} =G\x_H(\ms\oplus\{\mu\})\iso G\x_H\ms.$

A section, $s$, of $G/H\to G$ determines a trivialization of
$G\x_H\ms\iso T^*(G/H)$ via
\eqn{[s(q)h,p]_H&=&[s(q),h\cdot p]_H\\
&\to&(q,h\cdot p)\in G/H\x\ms} as every $g\in G$ has a unique
factorization $g=s(q)h$. So $h\cdot p$ represents a one form
$p_1\th^1+p_2\th^2=p_\ph d\ph+p_\th d\th\ll{plinkdef}$ at
$q=q(\ph,\th)$. Thus we need to find
$\{\th^i_{s_{N/S}(\ph,\th)}\}$. Starting with $s_N(\ph,\th)$, we find
that the local form for the left-invariant one-forms is
\eq{\left(\begin{array}{c}
\th_{s_N(\ph,\th)}^1 \\ \th^2_{s_N(\ph,\th)}\\ \th^3_{s_N(\ph,\th)}
\end{array}\right)
=H'(\ph)\left(\begin{array}{c}
-\sin\th d\ph\\d\th\\ \cos\th d\ph+d\ch\end{array}\right),}
where $H'(\ph)=G(0,0,\ph)$. Identifying
$p=\left(\begin{array}{c}p_1\\p_2\end {array}\right)$ with
$\left(\begin{array}{c}p_1\\p_2\\p_3=0\end {array}\right)\in \Bbb R^3$
we have,
$H'(-\ph)\cdot p=p'\ll{plink}$,
where
$p'=\left(\begin{array}{c}\frac{-p_\ph}{\sin\th
}\\p_\th\end{array}\right)$.
Given $(p_\th(q),p_\ph(q))$ we find the corresponding element of
$G\x_H\ms$, using either $s_N$ or $s_S$, is
$[g(\ph,\th,0),p']_H$.
We can now give (\ref{classlink}) explicitly; setting $p'_3=\mu$ so we
are identifying $G\x_H\ms$ with $G\x_H(\ms\x\{\mu\})$, we have
\[\nn\bar J_\mu(p_\th(q),p_\ph(q))=(G(\ph,\th,0)\cdot p',G(\ph,\th,0)\cdot
a_0)\hspace{7.2cm}\nn\]
\eq{\hspace{2.6cm}=\left(\left(\begin{array}{c}-p_\ph\cos\ph\cot\th
-p_\th\sin\ph+\mu\sin\th\cos\ph\\
-p_\ph\sin\ph\cot\th +p_\th\cos\ph+\mu\sin\th\sin\ph\\
p_\ph+\mu\cos\th\end{array}\right),\left(\begin{array}{c}\sin\th\cos\ph\\
\sin\th\sin\ph\\
\cos\th\end{array}\right)\right)\ll{observables}.}

Returning to the actual representations of \calG themselves, we really
require the derived representations. Let $\{u^i\}$ be the canonical
basis for $\Bbb R^3$. Using (\ref{derivrep}) and setting $\hat
q^i=\hbar d\pi^\mu(0,u^i)$, immediately we find for
$\ps^{N/S}\in L^2(M_{N/S},\Bbb C)$,
\eq{\hat q^i\ps^{N/S}(\ph,\th)=q^i\ps^{N/S}(\ph,\th)\ll{qaction},}
where $q(\ph,\th)=G(\ph,\th,0)\cdot a_0$. Finding the expression
$d\pi^\mu(X,0)$ requires some calculation. However, Landsman
\ci{LandII} has already done this for the very similar case of
$\calG={\rm SO(3)}\lx \Bbb R^3$ so we will not give the
details. Recalling that $\mu=n\hbar$, we find for $\hat
L^n_i=d\pi^\mu(a_i,0)$
\eq{(\hat L_1^n\ps^{N/S})(\ph,\th)=[\left(i\cos\ph\cot\th\cdd{}{\ph} +
i\sin\ph\cdd{}{\th} -
n\frac{\cos\ph}{\sin\th}(1\mp\cos\th)\right)\ps^{N/S}](\ph,\th)\ll{Jaction1};}

\eq{
(\hat L_2^n\ps^{N/S})(\ph,\th)=[\left(i\sin\ph\cot\th\cdd{}{\ph} -
i\cos\ph\cdd{}{\th} -
n\frac{\sin\ph}{\sin\th}(1\mp\cos\th)\right)\ps^{N/S}](\ph,\th);}
\eq{\hspace{-2.5in}
(\hat L_3^n\ps^{N/S})(\ph,\th)=[\left(-i\cdd{}{\ph} \mp
n\right)\ps^{N/S}](\ph,\th).\ll{Jaction3}} The term div in
(\ref{derivrep}) vanishes because if the vector field $X$ is complete,
div $ X=0$ by the $G$-invariance of $\nu$ (e.g., see \ci{AM}). Also
note that in the region the coordinate systems overlap, $\ps^N$ and
$\ps^S$ are related by (\ref{origpsilink}), namely
\eq{\ps^S(\ph,\th)=e^{2in\ph}\ps^N(\ph,\th)\ll{psilink}.}
The action of the generators of {\rm SU(2)}, detailed above, agree with
those given by Landsman \ci{LandII} for {\rm SO(3)} except that here
half-integer
values of $n$ are allowed, which reflects the fact that we are using
SU(2) rather than SO(3).

We can now give the quantization explicitly. Using (\ref{observables})
together with (\ref{theqzmap}), we have
\eq{\left.\begin{array}{rcl}-p_\ph\cos\ph\cot\th
-p_\th\sin\ph+\mu\sin\th\cos\ph
&\to&\hbar\hat L_1^n\\ &&\\ -p_\ph\sin\ph\cot\th
+p_\th\cos\ph+\mu\sin\th\sin\ph
&\to&\hbar\hat L_2^n\\&&\\ p_\ph +\mu\cos\th&\to&\hbar\hat L_3^n\\&&\\
\sin\th\cos\ph&\to&\hat q^1\\&&\\
\sin\th\sin\ph&\to&\hat q^2\\&&\\
\cos\th&\to&\hat q^3\end{array}\right\}\ll{GQqz}}
where the action of $\{\hat q^i\}$ and $\{\hat L_i\}$ on the
respective coordinate patches is given in (\ref{qaction}) and
(\ref{Jaction1}) - (\ref{Jaction3}). Note that this is the
quantization obtained when $\pi^\mu$ is restricted to \calG and thus
(\ref{GQqz}) does not give the quantum operators for all the observables
that could be quantized. It is, of course, straightforward to
calculate from (\ref{derivrep}) the quantum operators corresponding to
these other observables but, for simplicity, we have just restricted
ourselves to the ones corresponding to \calG via $\hat J_\mu$.

\ssect{Conclusion}
We have now achieved our aim of matching a preferred class of
observables with their quantum operator counterparts in a way which
satisfies the quantum conditions (Qi) to (Qiii) given in section
\ref{intro}, i.e., eq. (\ref{theqzmap}). Although having used the
method of geometric quantization we have managed to cast our results
for the quantum operators in the language of representations rather
than the form, given in (\ref{fhataction}), which is usually generated
by the geometric quantization approach. Thus, our results can be
considered to be a generalisation of Isham's \ci{Isham} approach ({\em
modified} in view of the results of section \ref{homspaces}) in that,
firstly, they are applicable in the case of a non-homogeneous
configuration space; and secondly, we now have a representation of Aut
$N\lx C^\infty_c(Q)$ rather than $G\lx V^*\subset{\rm Aut\ }N\lx
C^\infty_c(Q)$ and we thus have a correspondingly larger class of
physical observables that can be quantized.

Finally, for a particle in an external Yang-Mills field, the r\^ole of
the connection \al is now transparent. Note that we can regard \al to
be given by the classical Hamiltonian $H_0$ on $T^*N$ since $H_0$ implicitly
gives the metric on $N$ which then determines $\al$. Firstly, the
obvious r\^ole of the connection is in the quantum Hamiltonian
$H_\hbar$. Turning to the quantizing map $Q_\hbar$, however, we see
that this map is {\em independent} of the connection. This follows
since, for the symplectic leaves of $(T^*N)/H$, we could, given
$\calO_\mu\in\hs$, write the corresponding symplectic leaf as
$(J^{-1}(\calO_\mu))/H$ which is defined without recourse to the
connection. This is the reduced phase space of the particle. Similarly
both the map $\hat J_\mu$ and the derived representation $d\pi^\mu$
are defined without regard to the connection. (Recall that right hand
side of (\ref{derivrep}) is independent of the connection used.) Thus,
as claimed, the quantizing map is independent of the connection and, in fact,
there is only one set of quantum operators, labelled by $\calL({\rm
Aut\ }N\lx C^\infty_C(Q))$. Where the connection comes in, is that it
allows the `external' and `internal' classical variables of the
particle to be explicitly identified, i.e., it determines the local
form of $\hat J_\mu$ given in (\ref{genobservables}) where the
$(q^\al,p_\be)$ are considered as `external' variables and \nuu
represents the `internal' variables. Thus, the
connection determines the way in which the quantum operators are
interpreted at a physical level.

\end{document}